\pdfoutput=1
\documentclass[11pt,twoside,a4paper,cmspaper,final,collab]{cms-tdr}
\def\svnVersion{4fd92cb}\def\svnDate{2025/04/23}\def\cmsCernNoTag{CERN-EP-2024-269}\def\cmsCernDate{\today}\def\cmsMessage{Published in the European Physical Journal C as \href{http://dx.doi.org/10.1140/epjc/s10052-025-14097-x}{\doi{10.1140/epjc/s10052-025-14097-x}.}}
\begin{document}\cmsNoteHeader{MLG-24-001}

\newlength\cmsFigWidth
\ifthenelse{\boolean{cms@external}}{\setlength\cmsFigWidth{0.45\textwidth}}{\setlength\cmsFigWidth{0.49\textwidth}}
\ifthenelse{\boolean{cms@external}}{\providecommand{\cmsLeft}{upper\xspace}}{\providecommand{\cmsLeft}{left\xspace}}
\ifthenelse{\boolean{cms@external}}{\providecommand{\cmsRight}{lower\xspace}}{\providecommand{\cmsRight}{right\xspace}}

\newcommand{\pp}{\ensuremath{\Pp\Pp}\xspace}
\newcommand{\mt}{\ensuremath{m_{\PQt}}\xspace}
\newcommand{\hdamp}{\ensuremath{h_{\text{damp}}}\xspace}
\newcommand{\mtmc}{\ensuremath{m_{\PQt}^{\text{MC}}}\xspace}
\newcommand{\wMCi}{\ensuremath{w^{\text{MC}}_i}\xspace}

\newcommand{\rb}{\ensuremath{r_{\PQb}}\xspace}
\newcommand{\xb}{\ensuremath{x_{\PQb}}\xspace}
\newcommand{\ptB}{\ensuremath{\pt^{\PB}}\xspace}
\newcommand{\mb}{\ensuremath{m_{\PQb}}\xspace}
\newcommand{\mB}{\ensuremath{m_{\PQB}}\xspace}
\newcommand{\mW}{\ensuremath{m_{\PW}}\xspace}

\newcommand{\MINNLO}{MiNNLO\xspace}
\newcommand{\MINNLOPS}{MiNNLO\textsubscript{PS}\xspace}
\newcommand{\HVQ}{\textsc{hvq}\xspace}

\newcommand{\Njet}{\ensuremath{N_{\text{jet}}}\xspace}

\newcommand{\pone}{\ensuremath{p_1(x)}\xspace}
\newcommand{\ptwo}{\ensuremath{p_2(x)}\xspace}
\newcommand{\ponetwo}{\ensuremath{\pone/\ptwo}\xspace}

\newcommand{\pttt}{\ensuremath{\pt(\ttbar)}\xspace}
\newcommand{\etatt}{\ensuremath{\eta(\ttbar)}\xspace}

\newcommand{\chisq}{\ensuremath{\chi^2}\xspace}
\newcommand{\ndf}{\ensuremath{\text{NDF}}\xspace}

\cmsNoteHeader{MLG-24-001}
\title{Reweighting simulated events using machine-learning techniques in the CMS experiment}

\date{\today}

\abstract{Data analyses in particle physics rely on an accurate simulation of particle collisions and a detailed simulation of detector effects to extract physics knowledge from the recorded data. Event generators together with a \GEANT-based simulation of the detectors are used to produce large samples of simulated events for analysis by the LHC experiments. These simulations come at a high computational cost, where the detector simulation and reconstruction algorithms have the largest CPU demands. This article describes how machine-learning (ML) techniques are used to reweight simulated samples obtained with a given set of parameters to samples with different parameters or samples obtained from entirely different simulation programs. The ML reweighting method avoids the need for simulating the detector response multiple times by incorporating the relevant information in a single sample through event weights. Results are presented for reweighting to model variations and higher-order calculations in simulated top quark pair production at the LHC. This ML-based reweighting is an important element of the future computing model of the CMS experiment and will facilitate precision measurements at the High-Luminosity LHC.}

\hypersetup{%
pdfauthor={CMS Collaboration},%
pdftitle={Reweighting simulated events using machine-learning techniques in the CMS experiment},%
pdfsubject={CMS},%
pdfkeywords={CMS, machine learning, simulation, reweighting}}

{\tolerance=800
\maketitle}

\section{Introduction}

In particle physics, Monte Carlo (MC) event generators, paired with a
detailed simulation of the experimental apparatus, are ubiquitous.
The resulting MC simulations are crucial for simulating the detector response,
estimating the acceptance and reconstruction efficiency, and predicting signal and background
contributions from different processes expected in data.
These are indispensable for obtaining measurements corrected for detector effects
that can be directly compared to theoretical predictions.
In addition, MC simulations can be used for direct comparisons with collision data to extract
parameters of the underlying theory~\cite{Buckley:2009bj}.

Large MC samples with billions of events are needed
to achieve the physics programme goals of experiments at the CERN LHC, such that the statistical precision of MC samples has a small
effect on the total uncertainty of data analyses.
The generation and simulation of these MC samples come at a significant computational cost, where the
simulation of the detector response and the event reconstruction take more than 75\% of the
total CPU resources~\cite{CERN-LHCC-2022-005, Software:2815292}. In addition, the computing resources needed for producing
MC samples are expected to increase by factors of ten or more~\cite{HEPSoftwareFoundation:2017ggl}
at the High-Luminosity LHC (HL-LHC)~\cite{HEPSoftwareFoundation:2017ggl, Apollinari:2284929}.
The computing needs of the particle physics
programme at the HL-LHC exceed those that can be met by scaling up computing facilities.
Current estimates assume that 160 billion fully simulated and reconstructed MC events have to be
produced per year after the start of the HL-LHC. This number may become larger by up to 30\%
because of events with negative weights in next-to-leading order (NLO) and next-to-NLO (NNLO) simulations 
in quantum chromodynamics (QCD), which reduce the statistical precision of the MC samples.
Research and development in several
areas are ongoing to meet the strategic goal of minimizing CPU and storage requirements to enable the
LHC experiments to make full use of the physics potential offered by the large amount of data expected from the
HL-LHC.

An important part of these efforts is a reduction in the storage size and number of MC samples needed for
data analyses without compromising the precision and accuracy of analyses.
For the evaluation of systematic uncertainties connected to approximations made in the MC event generators,
additional samples are typically created. These samples are generated with fewer events than the nominal samples
because of limited computing resources. However, the smaller size of these samples can become a
limiting factor in precision analyses even with existing data sets.
For example, in a recent measurement of the top quark-antiquark (\ttbar) pair production cross section in
proton-proton (\pp) collisions at a centre-of-mass energy $\sqrt{s}=13\TeV$, a simultaneous fit
of the \ttbar cross section and the mass parameter of the top quark in the MC simulation \mtmc
is performed~\cite{CMS:2018fks}. The dominant contribution to the uncertainty in \mtmc arises from the
limited statistical precision of the MC samples used for the systematic uncertainty estimation.
While a regularization procedure might be considered to smooth out fluctuations, such an approach
harbours the risk of weakening genuine physical effects and does not prevent the need to simulate the
detector response for additional samples.

In this article, a method is introduced that allows for reweighting MC samples at generator level to represent many relevant
aspects from various simulation programs or generated with different parameter values.
The reweighting of the entire event is obtained using machine-learning (ML) algorithms, where the output can be stored
and applied on a centrally produced MC sample generated with a set of nominal parameter values.
In this way, the detailed detector simulation is needed only for one MC sample, which lowers the computational cost.
In addition, this approach allows the determination of modelling uncertainties with higher precision
because of the increased statistical accuracy of the nominal sample compared to the
smaller samples obtained with modelling variations.
While the generation of other samples than the nominal one is still needed for the training
of the ML model, the detector simulation and reconstruction algorithms do not have to be run
on these samples, since the reweighting is done on kinematic variables available at generator level.
This saves a considerable amount of CPU resources. In addition, the ML model
takes negligible storage space compared to samples with millions of events, such that this approach
helps to reduce the data storage needs of particle physics experiments.
The user can seamlessly incorporate the trained ML model into the analysis workflow at any stage.

{\tolerance=1800
Reweighting can be described in terms of multi-dimensional densities $p_i$. 
The conditional densities \pone and \ptwo describe two densities in a
set of given quantities of interest $x$.
The density ratio estimate \ponetwo is applied as a weight to the density \ptwo to obtain a conditional density matching \pone.
In standard reweighting approaches, the density ratio estimate is calculated in intervals of $x$, 
often realised by binning the two densities in one- or two-dimensional histograms. 
The density \ptwo is then multiplied by the approximate density ratio estimate in the same intervals of $x$ 
to arrive at a reweighted conditional density which should resemble \pone. 
The result is sensitive to the chosen size of the intervals and the dimension of $x$ is in practice
not larger than two because higher dimensionalities necessitate very large sample sizes.
Instead, reweighting techniques based on ML do not have these limitations.
The dimension of $x$ can be much larger than in standard reweighting methods and the training can be performed by
continuously sampling from the initial and target distributions, such that inaccuracies due to
binning effects are avoided and correlations between the elements of $x$ are accurately reproduced.
In addition, ML techniques allow for an interpolation of the
reweighted densities in the parameters, resulting in a continuous dependence
on the parameters in the simulation of collision events.
Several ML methods have been developed in the context of particle physics for reweighting simulations,
including boosted decision trees to derive event weights~\cite{Rogozhnikov:2016bdp},
neural networks (NNs) to learn the likelihood ratio between different
simulations~\cite{Cranmer:2015bka, Andreassen:2019nnm, Nachman:2021opi},
or input convex neural networks~\cite{pmlr-v70-amos17b} to calibrate simulated events~\cite{Pollard:2021fqv}.
Normalizing flows~\cite{normflows1, normflows2} can be used to learn a mapping between the initial
and target distributions, \eg to calibrate simulation to match data~\cite{Golling:2022nkl},
but also to learn the conditional probability distribution of the initial data,
from which events following a different conditional distribution can be sampled~\cite{Raine:2022hht,
Hallin:2021wme, Algren:2023qnb}. Reweighting techniques based on ML methods can also
enhance the precision of generative adversarial networks~\cite{Diefenbacher:2020rna}.
\par}

The ``deep NN using classification for tuning and reweighting (DCTR)''~\cite{Andreassen:2019nnm} method
can be applied to different scenarios.
In this article, the DCTR approach is used for reweighting simulated events of \ttbar production
to two modelling variations, important for the estimation of systematic uncertainties in the CMS experiment~\cite{CMS:2008xjf}.
In addition, the reweighting of an NLO simulation to an
NNLO simulation of \ttbar production is presented.

{\tolerance=1200
The paper is organized as follows. In Section~\ref{sec:DCTR}, a brief description of the ML reweighting method is given.
In Section~\ref{sec:Hdamp}, the discrete reweighting is presented 
of a parameter that changes the amount of radiation from the parton shower (PS) in events generated with
\POWHEG~v2~\cite{Alioli:2010xd, Frixione:2007nw, Frixione:2007vw}.
The continuous reweighting of a parameter affecting the fragmentation of \PQb quarks in \PYTHIA8~\cite{Sjostrand:2014zea} is described in Section~\ref{sec:Bfragm}.
The predictions obtained from the NLO \POWHEG \HVQ~\cite{Frixione:2007nw}
event generator are reweighted
to the predictions from the \POWHEG \MINNLO~\cite{Monni:2019whf, Monni:2020nks, Mazzitelli:2020jio, Mazzitelli:2021mmm} event generator at NNLO, both interfaced
with \PYTHIA8 for the simulation of the PS and hadronization, in Section~\ref{sec:MiNNLO}.
Finally, the implementation of the method within the CMS analysis and reconstruction software
is described in Section~\ref{sec:CMSSW}. The paper is summarized in Section~\ref{sec:conclsuions}.
\par}

{\tolerance=4000
\section{Deep neural network using classification for tuning and reweighting}\label{sec:DCTR}}

The DCTR approach is based on a deep NN to reweight MC samples obtained from
one simulation to resemble the features of a different simulation or a simulation with varied model parameters.
This is achieved by using the full kinematic and flavour information in the event~\cite{Andreassen:2019nnm}.
The method combines an ML architecture for including the particle information~\cite{Komiske:2018cqr}
with a parameterized classifier~\cite{Cranmer:2015bka, Baldi:2016fzo} to reweight one simulation
such that it resembles all relevant features of another.
We have chosen this approach for two main reasons.
First, it allows for full phase space reweighting, using all the event
information such as kinematic and particle type variables and enabling posterior projections onto any desired variable.
Second, this method allows for a continuous reweighting as a function of any parameter in the simulation,
which is exploited in the \PQb quark fragmentation study in Section~\ref{sec:Bfragm}.
A full phase space reweighting is important to achieve the best possible precision
when estimating modelling uncertainties from a change of parameters in the MC simulation.
Its significance becomes particularly pronounced in the context of reweighting
from a simulation at NLO to the one at NNLO accuracy.
In this scenario, where we have a three-particle phase space (\PQt, \PAQt, and one additional parton)
and a four-particle phase space (\PQt, \PAQt, and two additional partons),
the method remains effective for all observables derived from the \ttbar system
using information about the \PQt and \PAQt quarks, and the combined \ttbar system in the training.
This stands in contrast to standard reweighting methods, typically involving ratios in bins of two distributions,
which might be suboptimal for observables not included in the calculation of these ratios.

\subsection{The likelihood ratio trick}

The first ingredient of the full phase space reweighting technique is a prescription to derive event weights.
Consider two simulations describing the same phase space, denoted as $\Omega$, characterized by the
probability densities \pone and \ptwo, where the set of quantities $x\in\Omega$.
Assuming both densities have the same support, the function $w(x)=\ponetwo$ serves as
the ideal per-event weight for transforming the second simulation to match the first.
This weight function can be effectively approximated by training an ML classifier
to discern between the two simulations.
For example, an NN function $f(x)$ is trained using the binary cross-entropy loss
\begin{equation}\label{eq:L1}
    \text{loss}\big(f(x)\big) = -\sum_{i\in\alpha}\ln f(x_i)-\sum_{i\in\beta}\ln\big(1-f(x_i)\big),
\end{equation}
where $\alpha$ and $\beta$ represent sets of events from the two simulations,
\ie the two classes the binary cross-entropy loss is set up to distinguish between.
The likelihood ratio trick states that $f(x)/(1-f(x))\approx\ponetwo$~\cite{Andreassen:2019nnm}.
As a consequence, the weight function $w(x)$, which is the quantity of interest,
can be approximated from the network function $f(x)$ as $w(x)\approx f(x)/(1-f(x))$.
The benefit of parameterizing $f(x)$ as an NN is that ML algorithms 
can take into account the large dimensionality of $\Omega$ when constructing $f(x)$.
In the case of MC events with weights, for example from simulations at NNLO accuracy,
Eq.~\eqref{eq:L1} is modified to become
\ifthenelse{\boolean{cms@external}}{\begin{multline}\label{eq:L1w}
    \text{loss}\big(f(x)\big) = -\frac{1}{N}\sum_i\wMCi\Big(t_i\ln f(x_i) \\ +(1-t_i)\ln\big(1-f(x_i)\big)\Big),
\end{multline}}{\begin{equation}\label{eq:L1w}
    \text{loss}\big(f(x)\big) = -\frac{1}{N}\sum_i\wMCi\Big(t_i\ln f(x_i)+(1-t_i)\ln\big(1-f(x_i)\big)\Big),
\end{equation}}
where \wMCi is the MC weight for event $i$,
$t_i$ the true label of an event that is either 1 or 0 depending on whether the event belongs to class 1 or 2,
respectively, and $f(x_i)$ is the network function predicting the class of event $i$.
A discussion on negative event weights is provided later in Section~\ref{sec:MiNNLO}.

An important reweighting scenario is when the
two simulations have been obtained using the same simulation
program but with different parameter values $\theta$.
In this case, $\theta$ becomes the reweighting parameter.
For example, when modelling uncertainties are evaluated, one may want to
transform $p_{\theta}(x)$ into $p_{\theta+\delta\theta}(x)$. This is equivalent to reweighting
an MC sample obtained with the nominal value of a given parameter
to match a sample with a variation in this parameter, $\theta+\delta\theta$.
Additionally, the NN reweighting approximation in Eq.~\eqref{eq:L1} can be extended to the
continuous case by adding $\theta$ as a parameter to the
network function which becomes $f(x,\theta)$~\cite{Cranmer:2015bka, Baldi:2016fzo}.
In this case, the training data are generated with a uniform distribution over a range of
$\theta$ and Eq.~\eqref{eq:L1} becomes
\begin{equation}\label{eq:L2}
    \text{loss}\big(f(x,\theta)\big) = -\sum_{i\in\theta_0}\ln f(x_i,\theta)-\sum_{i\in\Theta}\ln\big(1-f(x_i,\theta)\big).
\end{equation}
Here, $x_i$ is a set of features sampled from a given simulation. 
The first sum runs over all sets of features in the simulation obtained with the nominal value
of the reweighting parameter $\theta_0$. 
The second sum runs over all sets of features from different simulations, 
where the set of these values $\theta$ is denoted by $\Theta$.
The NN is trained to distinguish between the sample generated with $\theta_0$ 
and a sample obtained with a set of values $\Theta$.

\subsection{Neural network architecture}
\label{sec:NN}

The final component of the DCTR method is an NN architecture that can
efficiently capture all the crucial aspects of the phase space.
We use the particle-flow network (PFN)~\cite{Komiske:2018cqr}, based on the deep sets
framework~\cite{ZaheerKRPSS17}, for this task.
The PFN is constructed to include the full event information, such as the particle four-momenta
and auxiliary information, like the particle type.
The reweighting parameter $\theta$ can also be used to parameterize the NN with the loss function
given in Eq.~\eqref{eq:L2}.
In the work presented here, we only use a small number of particle four-momenta and auxiliary 
information as input to the PFN, as described below.

The PFN is composed of two interconnected neural networks, $F$ and $\Phi$,
\begin{equation}
    f(p) = F\left(\sum_{i=1}^N\Phi(p_i)\right),
\end{equation}
where $p$ represents the set of all particles and $p_i$ are the properties of particle $i$ (momentum and type),
as well as the reweighting parameter $\theta$.
The two networks are integrated into a single model, by sequentially passing data through both networks,
with the output layer of $\Phi$ connected to the input layer of $F$.
The network $\Phi$ processes each particle individually, providing a per-particle internal (latent) representation.
The network $F$ takes the sum of these latent representations from all particles to produce an overall
event-level representation.
The dimension of the latent space is determined by the number of dimensions
used to represent the input data in a compressed form within the NN.
Specifically, the input particles are embedded in an $l$-dimensional latent space using $\Phi$,
while $F$ is an NN that maps $R^l\to R$.
The intermediate step involves creating an event-level representation by summing the
latent representations of all particles.
Finally, the output of the network $F$ is a softmax discriminant, which distinguishes
between the two classes. More details on the PFN and the setup used in this article
are given in Ref.~\cite{Komiske:2018cqr}.

We set up the NN $\Phi$ with two hidden layers, each with 100 nodes.
The classifier $F$ comprises three hidden layers and two output nodes for binary classification,
where each hidden layer consists of 100 nodes.
The latent space has dimension 128.
The ReLU~\cite{agarap2018deep} activation function is used for all layers, except for the classification
output that uses the softmax function. All models are implemented in \textsc{keras}~\cite{chollet2017keras} with the
\textsc{TensorFlow} back end~\cite{tensorflow2015-whitepaper}, passing the data sequentially through both networks.
The \textsc{adam}~\cite{kingma2017adam} optimization algorithm is used to
update the NN parameters during the training for minimizing the cross-entropy loss function for 100 epochs,
where each epoch represents a complete pass of the entire training data set through the network.
The training and validation loss functions measure the performance of the NN on training and validation data, respectively,
checking how well the NN predictions match the true values.
To prevent overfitting during training, the ``early stopping'' method is used to stop the training
process if the validation loss does not improve for a certain number of epochs.
For example, a patience value of 10 indicates that the training is stopped after 10 epochs without
improvement of the validation loss.
We use an early stopping with patience values of 10 for both the parton shower matching and \PQb quark fragmentation
trainings, while we use patience values of 30 for the \MINNLO training. The learning rate is set to 0.001 in all scenarios.
A hyperparameter optimization, specific for each case, is performed for the size of the training inputs and the batch size,
where the latter refers to the number of events in the training samples
processed simultaneously before the model parameters are updated. 
To select the best set of hyperparameters, the training is repeated with different choices of hyperparameters 
and the training giving the best reweighting result is chosen. We find that more complex tasks require a larger 
batch size, which also depends on the number of events provided as input to the NN.
Specific samples have been generated for each of the scenarios discussed in this paper.

\section{Parton shower matching uncertainties in simulated \texorpdfstring{\ttbar}{ttbar} production}
\label{sec:Hdamp}

The CMS experiment uses a setup of \POWHEG{}+\PYTHIA8 to generate
\ttbar MC samples that are used in physics analyses.
This is done by simulating the hard process of \ttbar production with the \HVQ program~\cite{Frixione:2007nw}
in the \POWHEG~v2~\cite{Frixione:2007vw, Alioli:2010xd} generator,
which is interfaced with \PYTHIA8.240~\cite{Sjostrand:2014zea}
for parton showering and hadronization.
In this configuration, \POWHEG simulates inclusive \ttbar production with matrix elements 
at NLO accuracy in QCD.
A resummation damping factor called \hdamp
controls the matching of the matrix elements with the PS. It changes the scale of the first
emission from the hard process and regulates the radiation at high transverse momentum \pt.
The \hdamp parameter enters the simulation in a damping function $D$, which reduces
the real contribution to the Sudakov form factor~\cite{Alioli:2008tz} and is given by
\begin{equation}
    D = \frac{\hdamp^2}{\pt^2+\hdamp^2}.
\end{equation}
Here, \pt refers to the transverse momentum of the top quark or antiquark and
$\hdamp=h\mt$, where $h$ is a real number and \mt is the top quark mass of $\mt=172.5\GeV$.
This value is consistent with the most precise result on the top quark mass $\mt=172.52\pm0.33\GeV$,
obtained by the combination~\cite{ATLAS:2024dxp} of ATLAS and CMS measurements.
The parameter \hdamp affects the kinematic properties of all particles in the event,
most notably the \pt of the \ttbar system \pttt and its pseudorapidity \etatt,
where the definition of the coordinate system is given in Ref.~\cite{CMS:2008xjf}.

Since the MC parameter \hdamp cannot be derived from first principles, it needs to be determined
from measurements. 
Such estimations result in a sizeable uncertainty in \hdamp~\cite{Sirunyan:2019dfx},
which can in turn translate into one of the leading systematic uncertainties in precision studies of the top quark.
Because the \hdamp parameter cannot be reweighted internally by MC generators, additional MC samples
obtained for different values of \hdamp are necessary to estimate the associated modelling uncertainty.
Therefore, the \hdamp parameter is an ideal case for a reweighting with the DCTR method.

\subsection{The \texorpdfstring{\hdamp}{hdamp} parameter in CMS}

\begin{figure}[!ht]
\centering
\includegraphics[width=\cmsFigWidth]{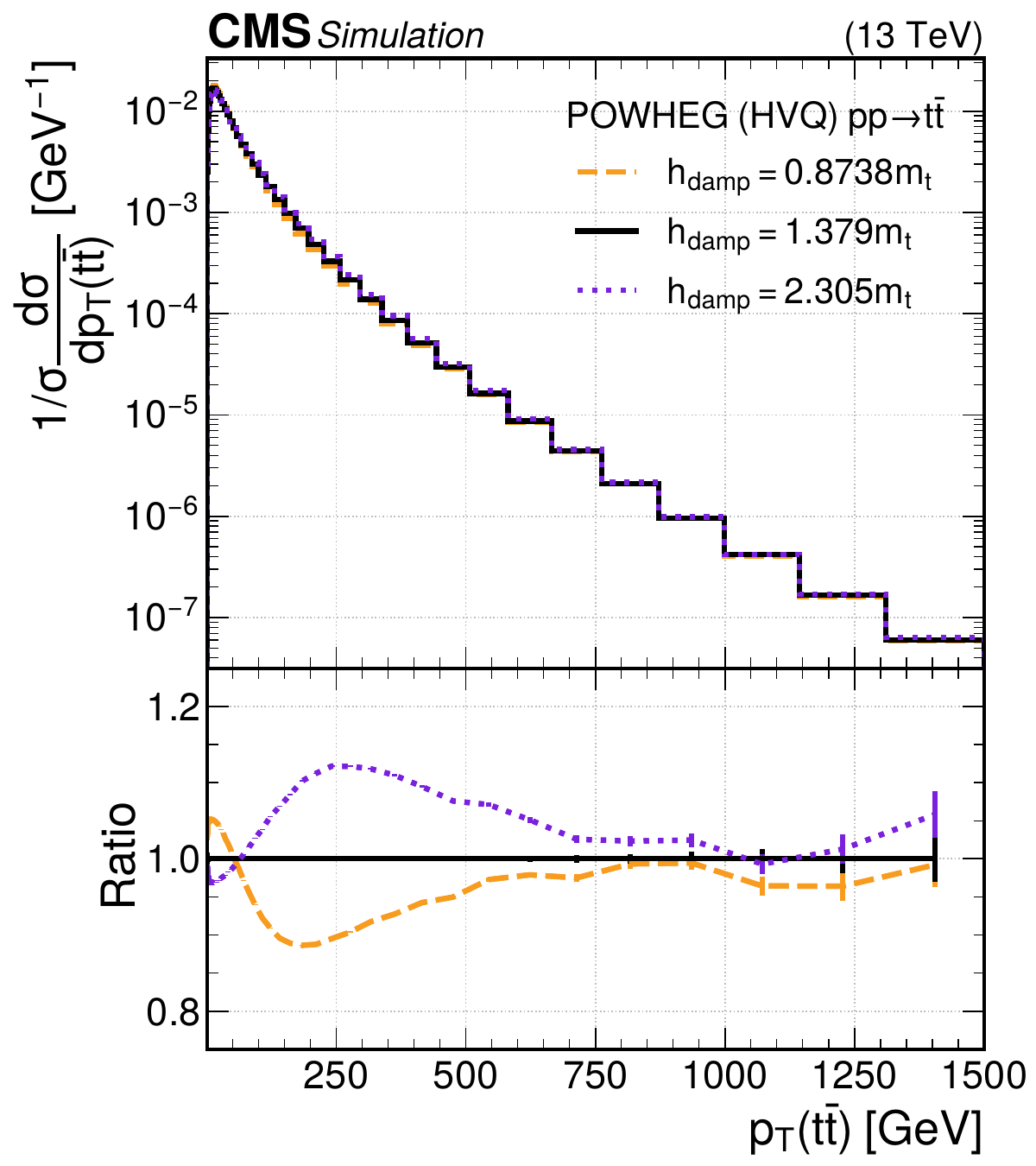}%
\hfill%
\includegraphics[width=\cmsFigWidth]{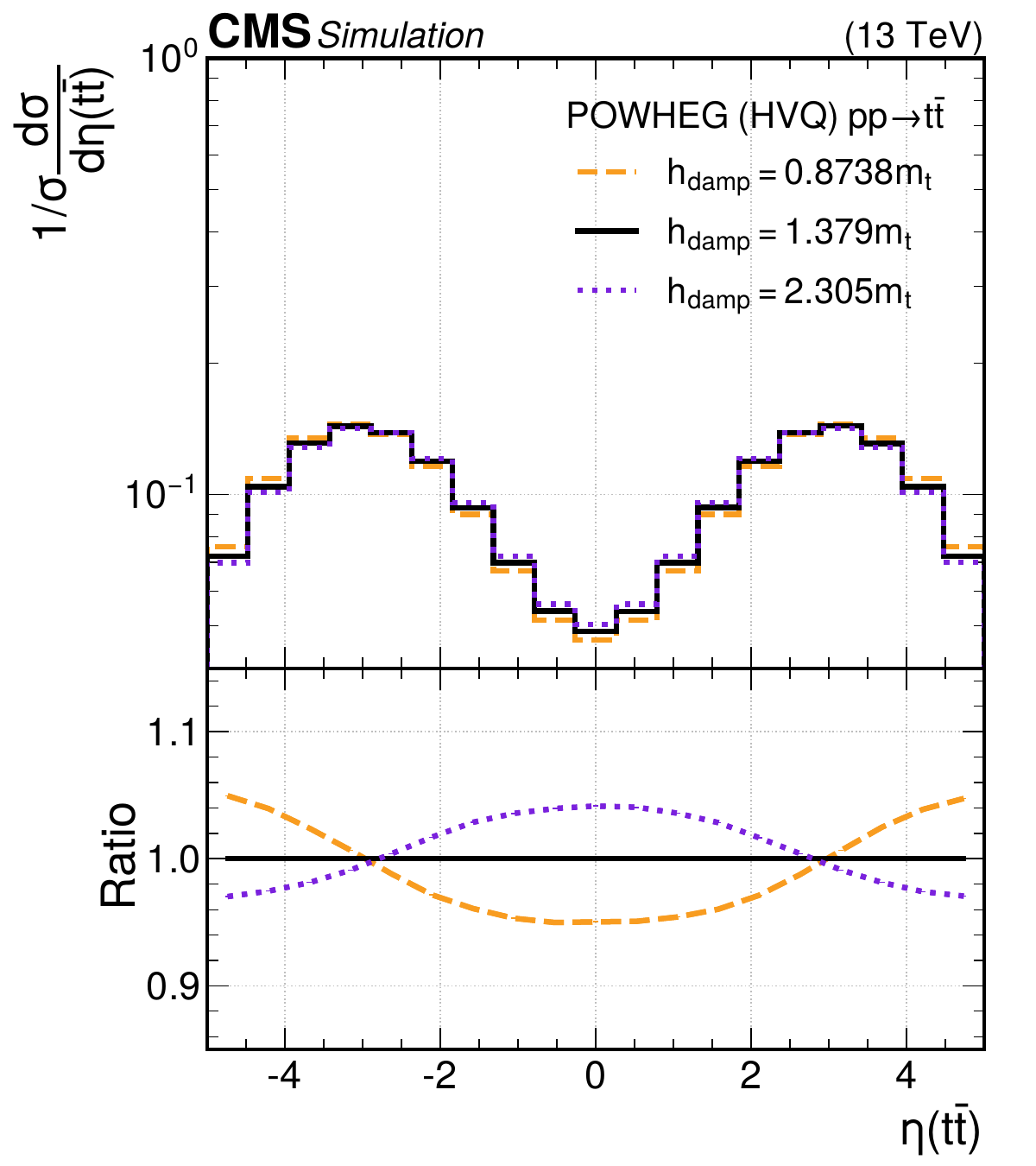}
\caption{The normalized differential cross section of \ttbar production in \pp collisions at 13\TeV
as a function of the \pt (\cmsLeft) and $\eta$ (\cmsRight) of the \ttbar system obtained with the \POWHEG program.
The standard setting of $\hdamp=1.379\,\mt$ (black solid lines) is compared to down (orange dashed lines) and
up (violet dotted lines) variations in \hdamp.
The ratios of the predictions with the \hdamp variations to the nominal one are shown in the lower panels.
The vertical bars, in the ratio panels, represent the statistical uncertainties in the MC samples.
}
\label{fig:BefRwgPt}
\end{figure}

{\tolerance=800
In the CMS experiment, two variations of the \hdamp parameter with respect
to its nominal value are considered to estimate the uncertainty associated with this parameter.
The nominal value of \hdamp is set to $1.379\,\mt$~\cite{Sirunyan:2019dfx}.
The up and down variations are $2.305\,\mt$ and $0.8738\,\mt$, obtained from a tune to data.
Figure~\ref{fig:BefRwgPt} shows the effect of these variations in \hdamp for distributions
in \pttt and \etatt.
Differences in the cross sections up to 10\% are observed in the resummation region of $\pttt\sim250\GeV$.
At higher \pttt, the effect is negligible and is expected to vanish for
$\pttt>1\TeV$~\cite{amoroso2020leshouches2019physics}, such that applying an NN weight of 1 is sufficient.
A difference of about 5\% is observed in \etatt.
\par}

In the CMS experiment, the samples generated with the two variations of \hdamp are produced
with fewer events, less than half the number of events in the nominal sample, where the latter could be of the order of several billion events.
The ML reweighting method studied in this paper presents a valuable
alternative, ensuring consistent MC statistical uncertainties from both the nominal and variation samples.
Since this approach increases the statistical precision of the samples with varied \hdamp parameters,
it promises to enhance the precision of future analyses.

\subsection{Training sample and NN parameters}

We train two different NN models to reweight the nominal \hdamp MC simulation to the two variations.
Each NN model uses a binary classification output to distinguish between the nominal and up/down variations.
The NN is trained using the parton-level information at the matrix element level, 
where the \PQt and \PAQt information is passed to the NN for training.
The particles are represented by their four-momenta and type, where the latter is encoded as a particle ID (PID) number.
The set of input variables per particle is given by (\pt, $y$, $\phi$, $m$, PID), where $y$ denotes the
rapidity, $\phi$ the azimuthal angle, and $m$ the mass.
The inputs to the NN are scaled to values of order $\mathcal{O}(1)$ before the training.
The parameter $m$ is divided by 244\GeV, which is found to be the maximum value of $m$ in the training sample,
and $\log(\pt/1\GeV)$ is used instead of \pt.
For each setting of the \hdamp parameter, 40 million events are generated.
Of the resulting 80 million events for the up and down variations in \hdamp,
75\% are used for training and 25\% for validation.
For both variations in \hdamp, we find that a batch size of 40\,000 events gives the best results.

\subsection{Results}

\begin{figure}[!ht]
\centering
\includegraphics[width=\cmsFigWidth]{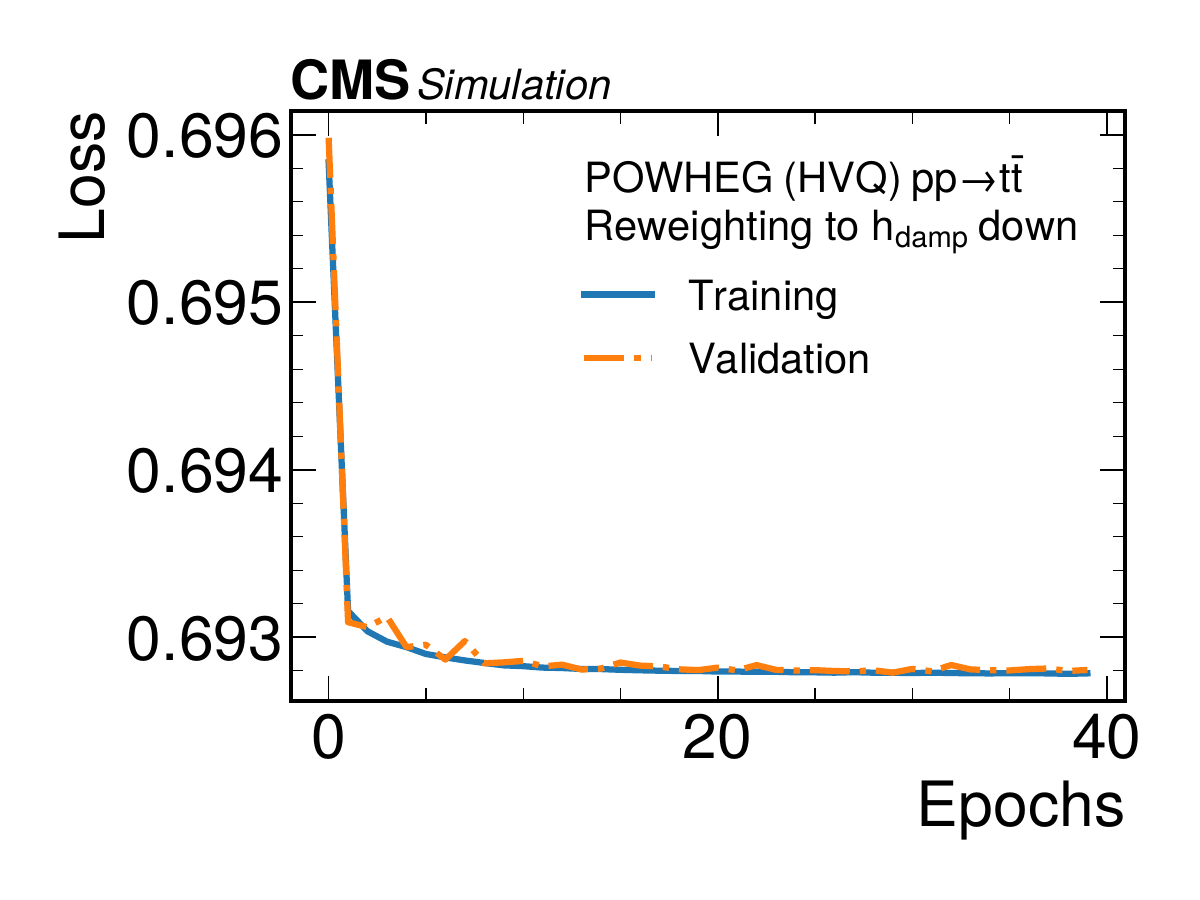}%
\hfill%
\includegraphics[width=\cmsFigWidth]{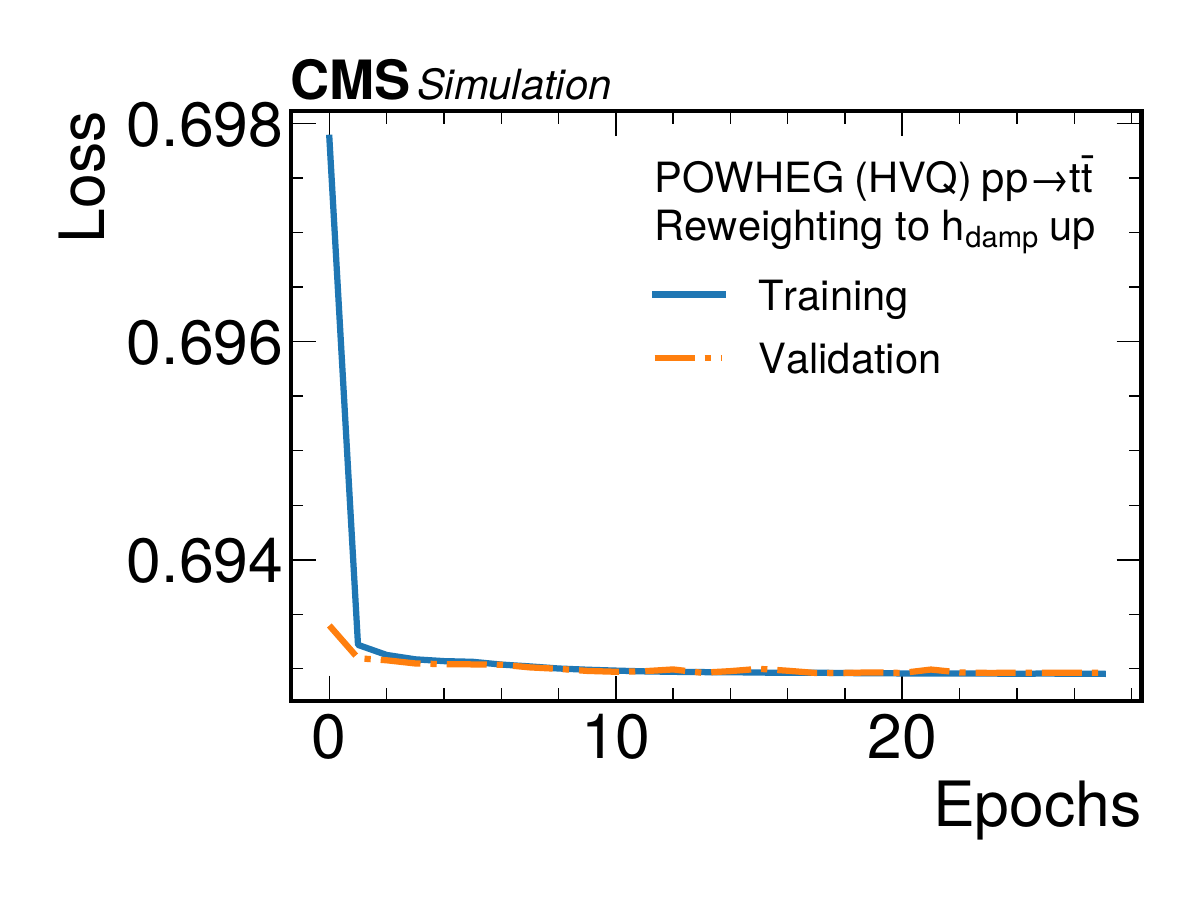}
\caption{The NN histories of the training for the \hdamp parameter reweighting. Shown are the
loss functions for the training data (blue solid line) and the validation data (orange dash-dotted line) for the
down (\cmsLeft) and up (\cmsRight) variations of \hdamp.
}
\label{fig:Loss}
\end{figure}

The performance of the training is evaluated by examining the values of the loss
functions on the training and validation data sets.
The graphs of the loss and validation loss functions are shown in Fig.~\ref{fig:Loss},
representing the NN models trained with the down (\cmsLeft) and up (\cmsRight) variations of \hdamp.
The NN has been trained for about 40 epochs for the model trained with the down variation of \hdamp
and 30 epochs for the model trained with the up variation of \hdamp.

\begin{figure*}[!tp]
\centering
\includegraphics[width=\cmsFigWidth]{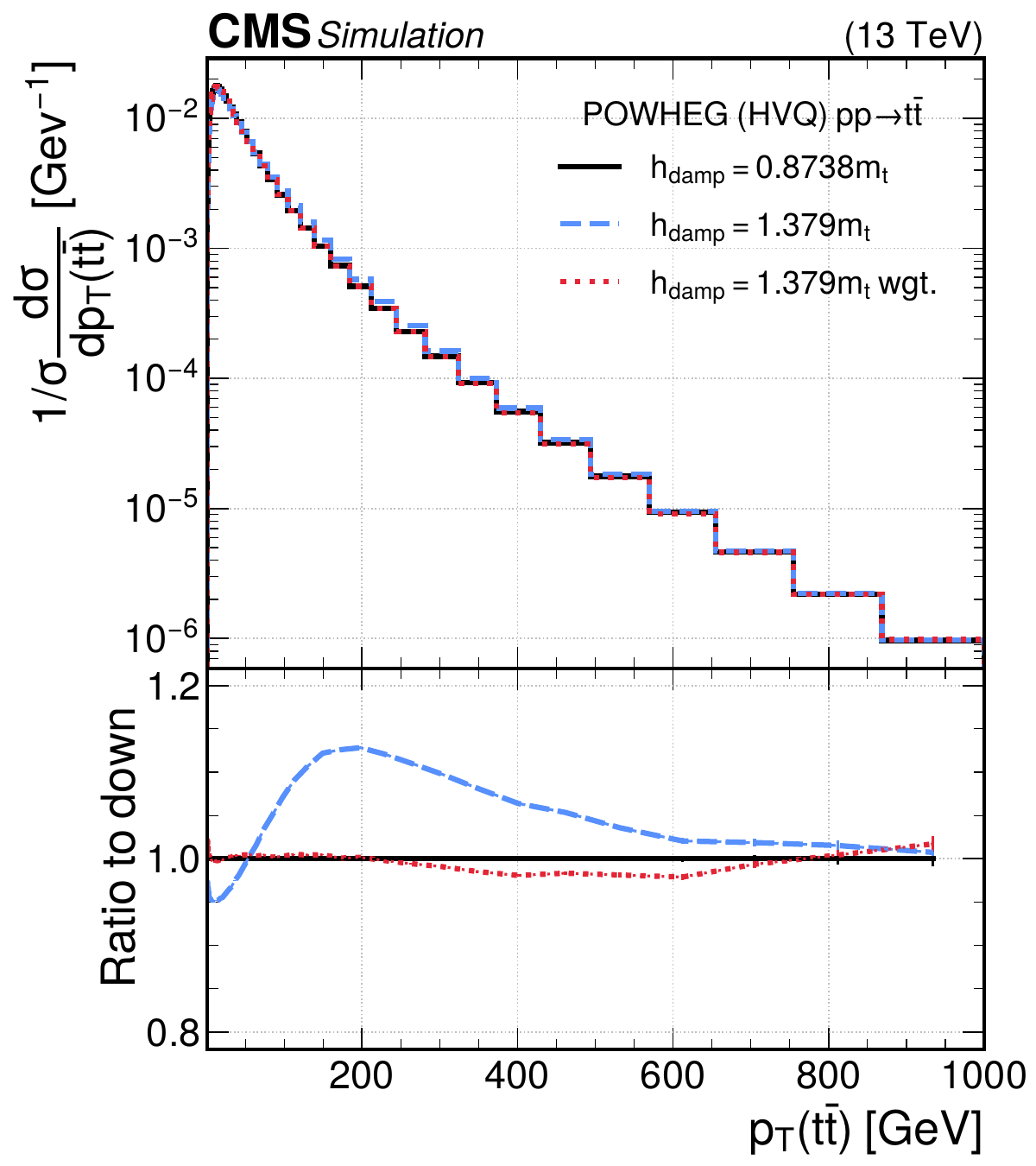}%
\hfill%
\includegraphics[width=\cmsFigWidth]{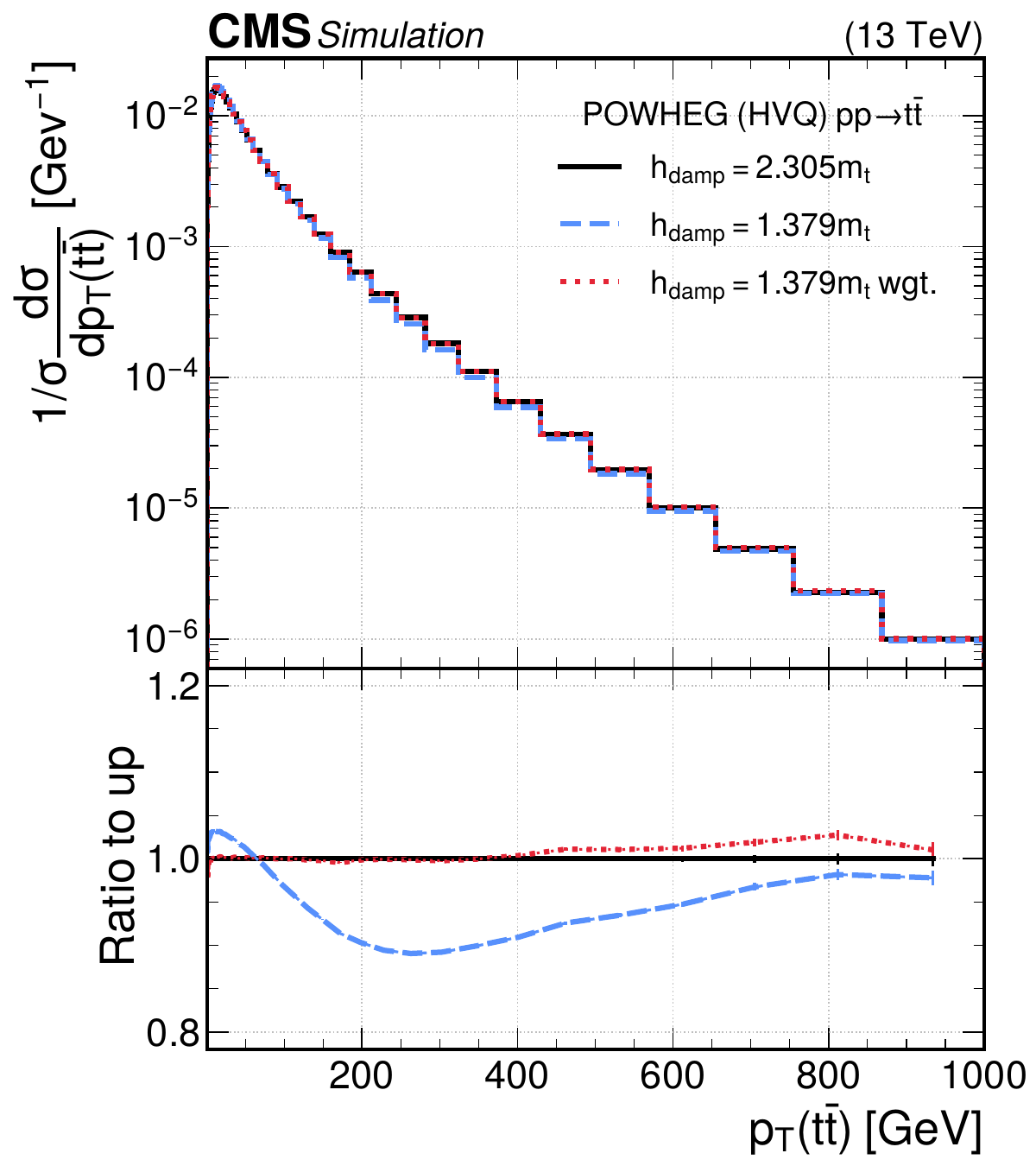} \\
\includegraphics[width=\cmsFigWidth]{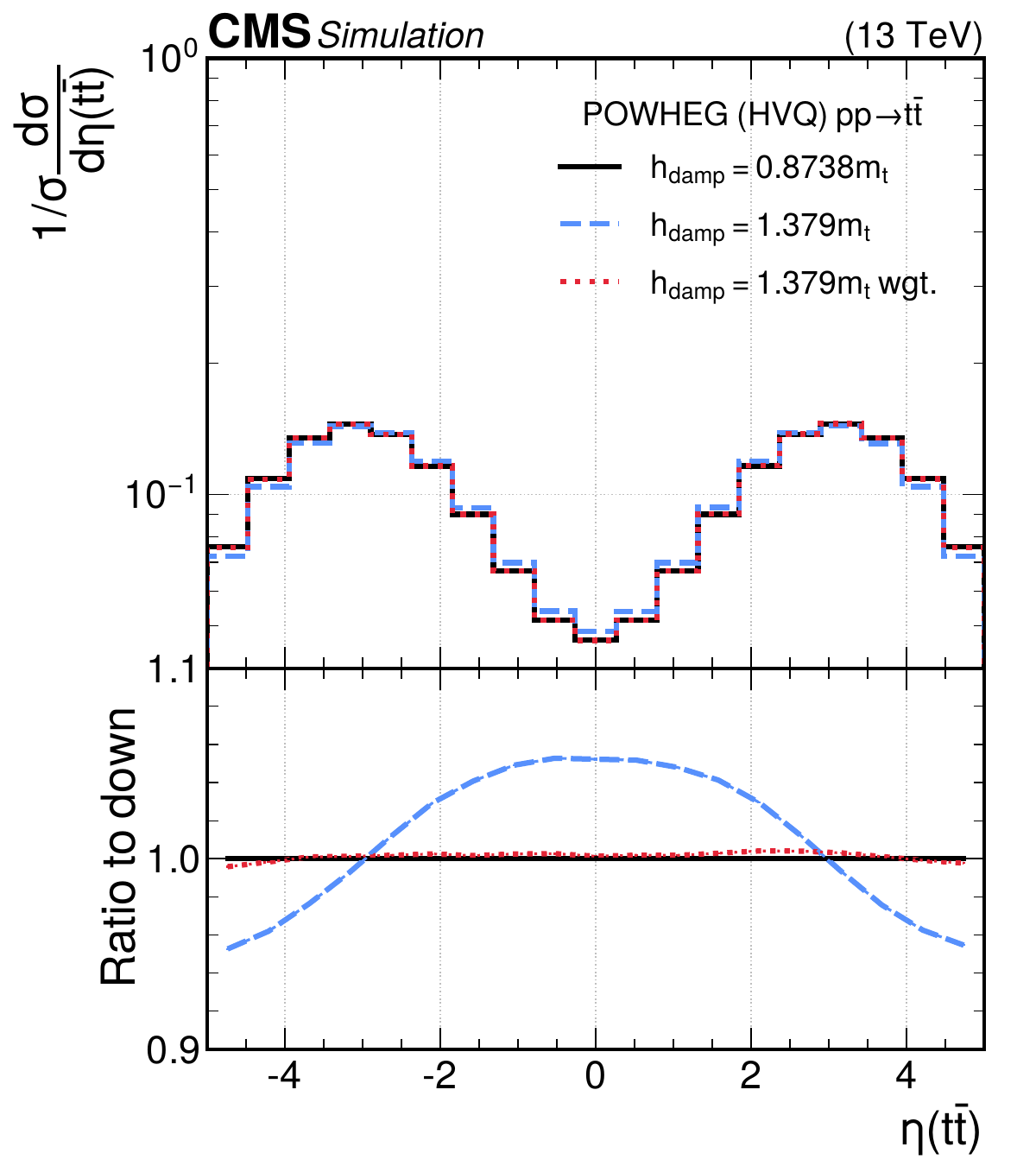}%
\hfill%
\includegraphics[width=\cmsFigWidth]{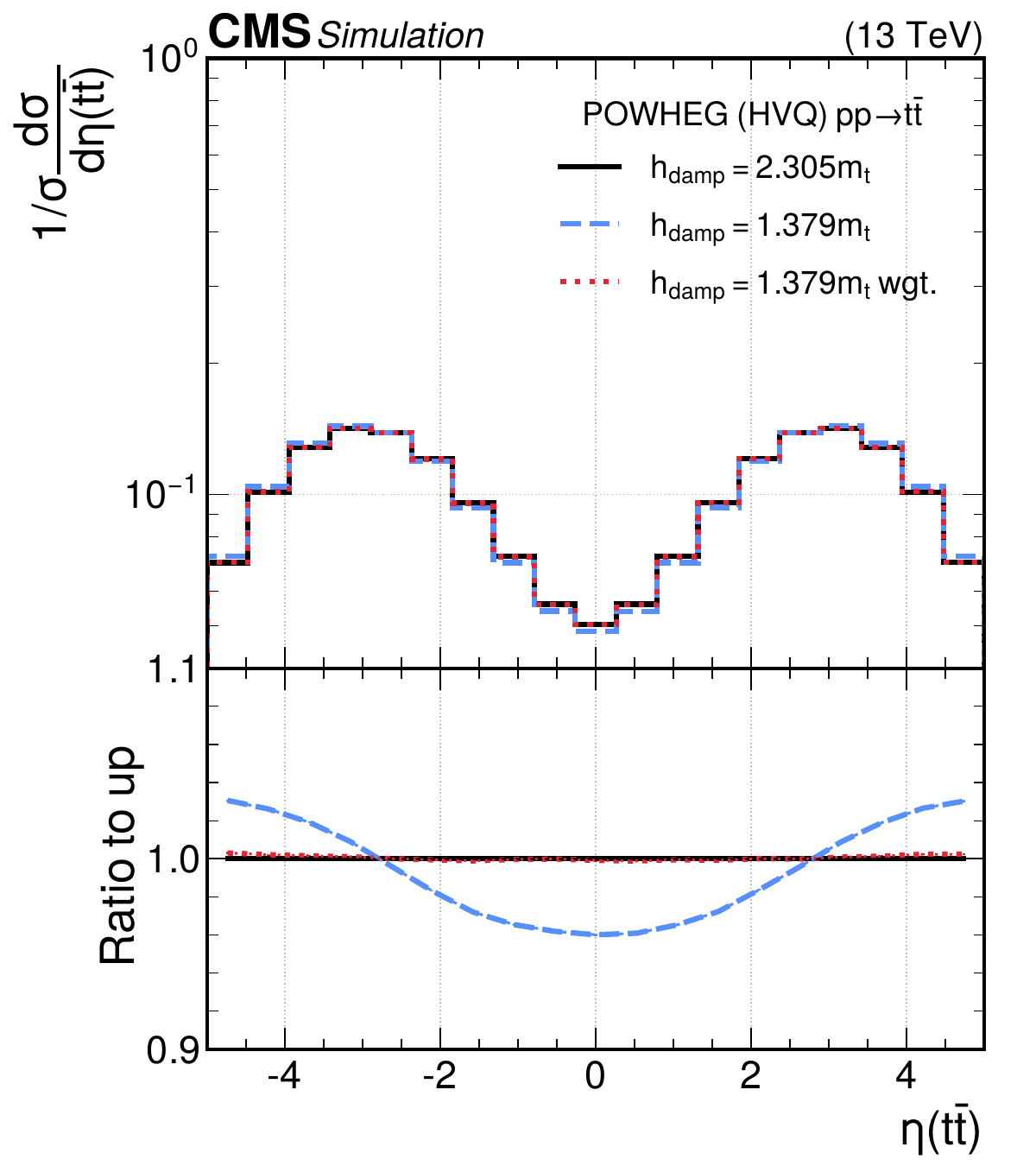}
\caption{The normalized differential cross section as a function of the \pt (upper) and $\eta$ (lower)
of the \ttbar system. The black solid line shows the predictions from the down (left) and up (right) variations in \hdamp,
and the blue dashed line presents the prediction from the nominal sample.
The red dotted line indicates the nominal sample reweighted to the down (left) and up (right) \hdamp variations
using the DCTR method.
The ratios to the samples with the target values of \hdamp are displayed in the lower panels,
together with their almost negligible statistical uncertainties (vertical error bars).
\label{fig:TopPair}
}
\end{figure*}

We test the performance of the reweighting using statistically independent test samples of 100 million events.
The weights are found to be close to unity with a standard deviation of about 0.1.
No weights smaller than 0.8 and larger than 1.2 are observed.
The weights from the trained model are then applied to reweight an MC sample with
$\hdamp=1.379\,\mt$ to the up and down variations with $\hdamp=2.305\,\mt$ and $0.8739\,\mt$, respectively.
The effect of the reweighting on the distributions in \pttt and \etatt
is shown in Fig.~\ref{fig:TopPair}.
The accuracy of the reweighting is quantified by the ratio to the nominal sample, shown below each distribution.
The target samples, \ie the samples generated with the variation of \hdamp,
and the reweighted ones agree within 1\% in \etatt.
Deviations of up to 2\% are observed for $\pttt\gtrsim400\GeV$.
Translating this deviation into an uncertainty, the precision of the reweighting
is comparable to the statistical precision of a sample with about 250 million events
taking into account branching fractions of the \ttbar system and typical analysis efficiencies.
The effect of the ML reweighting on the \pt and $y$ distributions of the additional parton
has also been checked, reaching the same precision as that of the \ttbar system.

\begin{figure}[!ht]
\centering
\includegraphics[width=\cmsFigWidth]{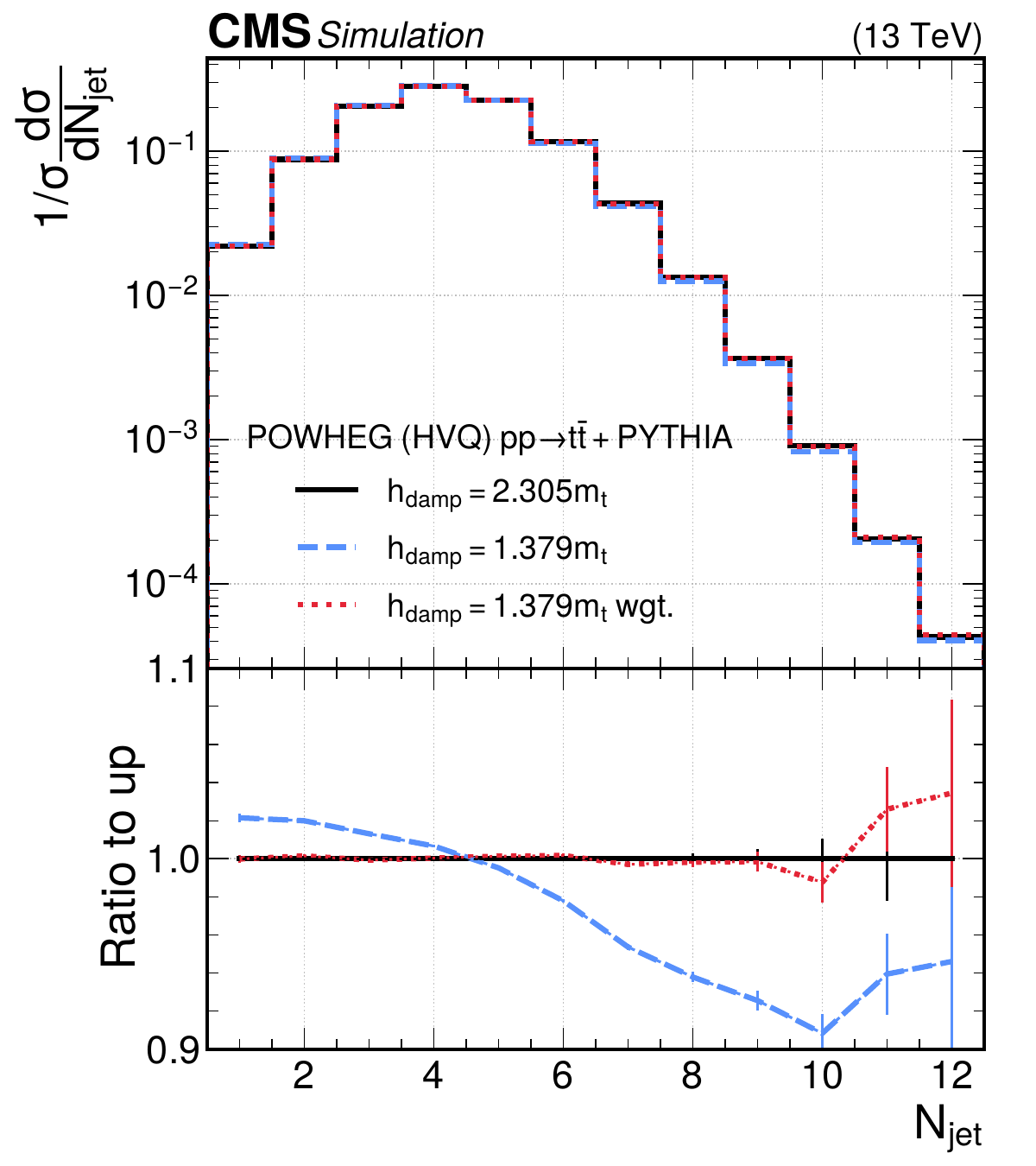}%
\hfill%
\includegraphics[width=\cmsFigWidth]{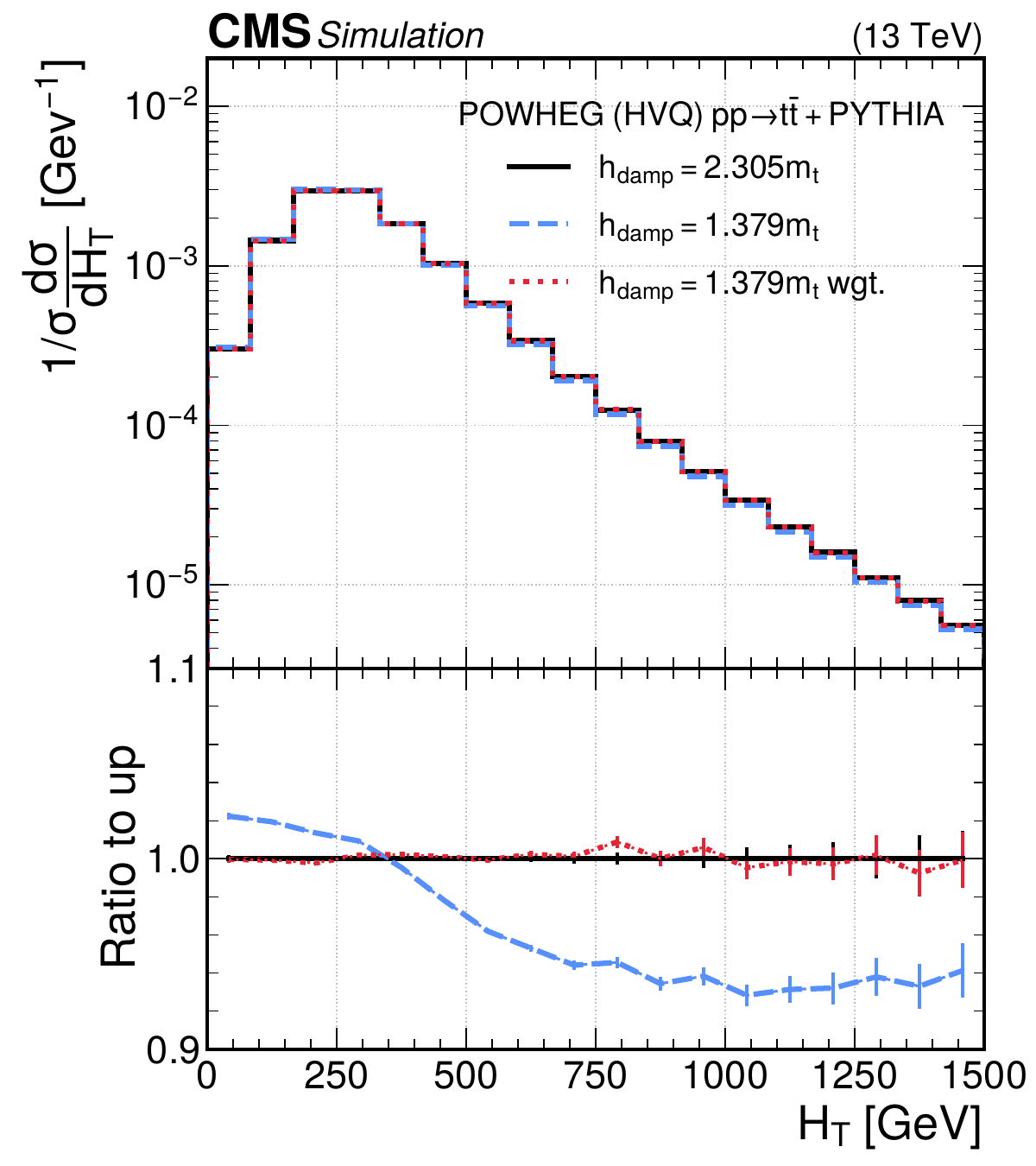}
\caption{The normalized differential cross section as a function of \Njet (\cmsLeft)
and \HT (\cmsRight). The black solid line shows the predictions from the up variation in \hdamp
and the blue dashed line presents the prediction from the nominal sample.
The red dotted line indicates the nominal sample reweighted to the \hdamp variation using the DCTR method.
The ratios to the target distributions are displayed in the pads below, where the vertical bars represent statistical uncertainties.
}
\label{fig:Njet}
\end{figure}

The accuracy of the reweighting is further tested on \ttbar events after the PS and
hadronization, where both effects have been simulated with \PYTHIA8.
This test is performed to verify that the model trained on parton-level information
works for events at the level of stable particles (particle level).
Results of the reweighting for distributions obtained at the particle level are shown in Fig.~\ref{fig:Njet}.
Distributions in the jet multiplicity \Njet and the scalar \pt sum of all jets in the event (\HT)
are presented for jets with $\pt>30\GeV$ and $\abs{\eta}<2.4$.
The jets are found using the anti-\kt jet algorithm~\cite{Cacciari:2008gp} with a distance parameter of $R=0.4$.
The target distributions obtained from samples generated with the up variation of \hdamp
agree with the reweighted ones within 1\% in both observables.

\subsection{Statistical uncertainty of the method}

\begin{figure}[!ht]
\centering
\includegraphics[width=\cmsFigWidth]{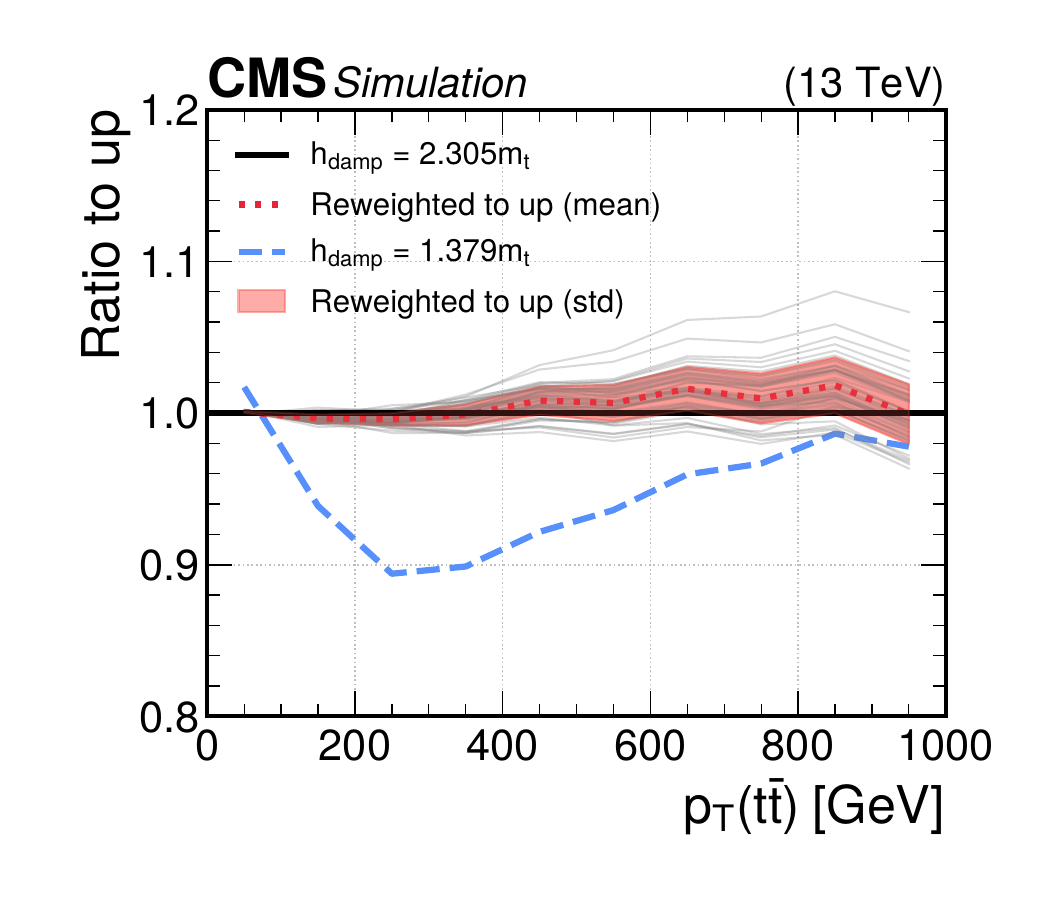}%
\hfill%
\includegraphics[width=\cmsFigWidth]{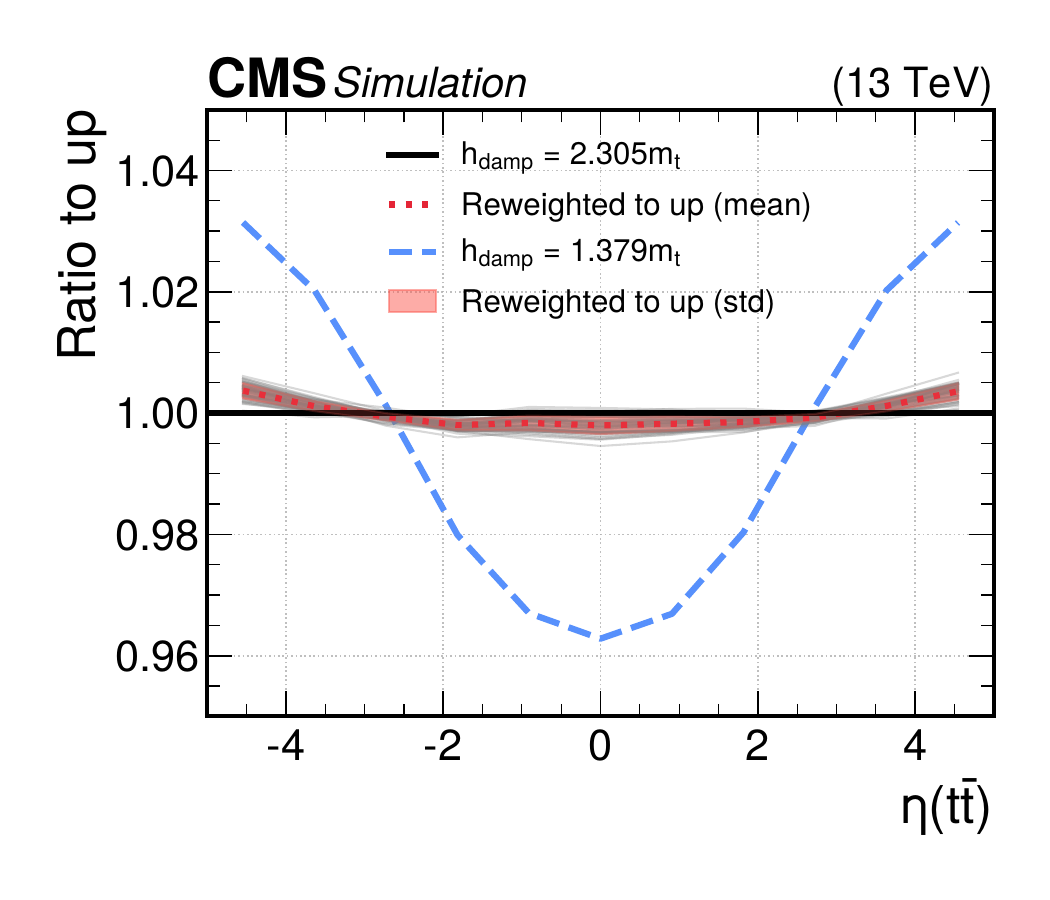}
\caption{Ratios between the \hdamp target distributions in \pttt (\cmsLeft) and \etatt (\cmsRight),
and 50 different reweightings (grey solid lines). The ratio to the target before the reweighting
is shown as a blue dashed line and the mean of the different reweightings as a red dotted line.
The red band represents the statistical uncertainty of the method obtained from the
standard deviation of the 50 reweighted samples.
\label{fig:Stat}
}
\end{figure}

{\tolerance=800
The statistical precision of the reweighting method is evaluated by repeating the training procedure 50
times~\cite{Efron1992}.
Each training uses 80 million events, randomly chosen from a sample of 200 million events.
The results from the 50 different trainings, expressed in ratios of reweighted to target distributions
in \pttt and \etatt, are presented in Fig.~\ref{fig:Stat}.
The average from the 50 reweighted distributions is shown together with the statistical uncertainty,
obtained from the standard deviation of these distributions. The statistical uncertainty
ranges from less than 0.5\% at small values of \pttt up to about 2\% at \pttt of about 900\GeV.
In \etatt, a statistical uncertainty of less than 1\% is found.
The target sample, generated with the up variation of \hdamp, and the average value of the 50 reweighted
samples are found to be compatible within the statistical uncertainty of the method, except 
for $\abs{\etatt}>4$ where we observe a difference between the two samples of 
between one and two standard deviations. This difference is smaller than 0.5\%, 
which is much less than the precision of our simulations and can be neglected.  
The statistical uncertainty of the method is of the same size as the deviation between the
target distribution and the nominal result in Fig.~\ref{fig:TopPair}.
\par}

\section{The \texorpdfstring{\PQb}{b} quark fragmentation}
\label{sec:Bfragm}

{\tolerance=800
A significant source of uncertainty in precision top quark studies is the
fragmentation of \PQb quarks into hadrons~\cite{CMS:2018fks, CMS:2023ebf, ATLAS:2024dxp}. 
The \PQb quark from the decay $\PQt\to\PW\PQb$
fragments into a \PQb hadron and several lighter hadrons.
This process is described in \PYTHIA by the Lund string model~\cite{Sjostrand:2014zea},
where the probability for the \PQb hadron to carry the momentum fraction $z$
of the \PQb quark momentum is given by the Lund--Bowler fragmentation function~\cite{bierlich2022comprehensive, Bowler:1981sb},
\begin{equation}
    f_{\PB}(z) = \frac{1}{z^{1+b\rb\mb^2}}(1-z)^a\exp\left(-\frac{b\mT^2}{z}\right).
\end{equation}}
The transverse mass of the \PQb hadron is given by $\mT^2=\mB^2+\Big(\ptB\Big)^2$,
where \mB and \ptB are the \PQb hadron mass and transverse momentum, respectively.
The free parameters $a$ and $b$ are treated as universal for all quarks.
Their values $a=0.68$ and $b=0.98\GeV^{-2}$ were obtained from a fit to data
sensitive to light-quark fragmentation in the Monash tune~\cite{Skands:2014pea},
such as charged-particle multiplicities and momentum fractions.
The Lund--Bowler parameter \rb and the \PQb quark mass \mb are specific to the fragmentation of \PQb quarks,
where the latter is set to 4.78\GeV~\cite{Skands:2014pea}.
In the Monash tune, which is the default for fragmentation in \PYTHIA8, $\rb=0.855$ is obtained.

Changing the \rb parameter in the event simulation affects certain observables sensitive
to the \PQb quark fragmentation. An important example is \xb, which
is defined as the normalized \PQb hadron energy fraction.
At the level of stable particles, it is given by~\cite{Corcella:2001hz, Cacciari:2002re}
\begin{equation}
    \xb = \frac{2p_{\PB}\cdot p_{\PQt}}{\mt^2}\frac{1}{1-\mW^2/\mt^2},
\label{eq:xb}
\end{equation}
where $p_{\PB}$ and $p_{\PQt}$ are the four-momenta of the \PQb hadron and the top quark, respectively,
and \mW is the \PW boson mass.
Changes in the value of \rb affect the distribution in \xb,
which can lead to efficiency differences in the identification of
\PQb jets and can result in sizeable systematic uncertainties in data analyses.

{\tolerance=800
In CMS, the default underlying event tune for 13\TeV analyses is the CP5 tune~\cite{Sirunyan:2019dfx}.
Besides other differences from the Monash tune, it uses a smaller value of the
strong coupling constant at the mass of the \PZ boson in the final-state shower, $\alpS(M_{\PZ})=0.118$,
than the Monash tune, which uses $\alpS(M_{\PZ})=0.1365$.
The \rb value is not changed in the CP5 tune.
Nevertheless, a re-derivation of the \rb parameter on top of the CP5
tune with \EE data from LEP results in \smash{$\rb=1.056\,^{+0.196}_{-0.200}$}.
In this paper, two NN models are constructed to reweight the sample generated with the CP5 tune
with $\rb=0.855$ to the nominal and up variation of the \rb parameter values of 1.056 and 1.252.
For the down variation, the generated value of $\rb=0.855$ is used, which almost coincides
with $\rb=0.856$ obtained from the reanalysis of LEP data.
In addition, a continuous reweighting is implemented using ten distinct values of the \rb parameter during training.
A single NN model is trained to reweight the generated sample to any arbitrary
value of \rb within the interval $[0.6,1.4]$, allowing for an interpolation among the ten \rb values.
\par}

\subsection{Training sample and NN parameters}

Approximately one million events for each \rb value are used in the discrete reweighting
for the training and validation.
In the case of the continuous reweighting, the NN is trained on 5 million events,
where 2.5 million events were generated with $\rb=0.855$ and
another 2.5 million events for ten different \rb values within the interval $[0.6,1.4]$,
\ie 250\,000 events were generated for each value of \rb.
In both scenarios, 90\% of the events are allocated for training and 10\% for validation.
The NN architecture is identical to the one described in Section~\ref{sec:NN}.
A batch size of 1000 is found to give optimal results.
The NN is trained using the particle-level information of the events produced with \PYTHIA8.
In this case, the input variables to the NN are \xb of \PQt and \PAQt, as given in Eq.~\eqref{eq:xb},
and the value of \rb serving as reweighting parameter.
While this specific scenario might be manageable with a standard one-dimensional reweighting approach, 
extending it to a continuous reweighting requires an ML reweighting approach.
The observable \xb comprises the relevant information from the event,
as it includes the four-momenta of the top quarks and the \PQb hadrons in the event.
The four-momenta of \PQt and \PAQt are taken from the last copy in the
\PYTHIA8 event record before the decay, and the first copy of the \PQb hadrons are used.
The input variable \rb is close to unity and the variable \xb is defined in the interval $[0,1]$.
However, while the energy of a \PQb hadron cannot exceed that of the parent bare \PQb quark,
this is no longer true in a real collision event, where energy in the hadronization can come
from other parts of the final state, such as the underlying event, which is particularly relevant at the LHC.
As a result, although most events fall within the range $\xb\in[0,1]$, a few values of \xb may exceed unity.

\subsection{Results}

\begin{figure*}[!tp]
\centering
\includegraphics[width=\cmsFigWidth]{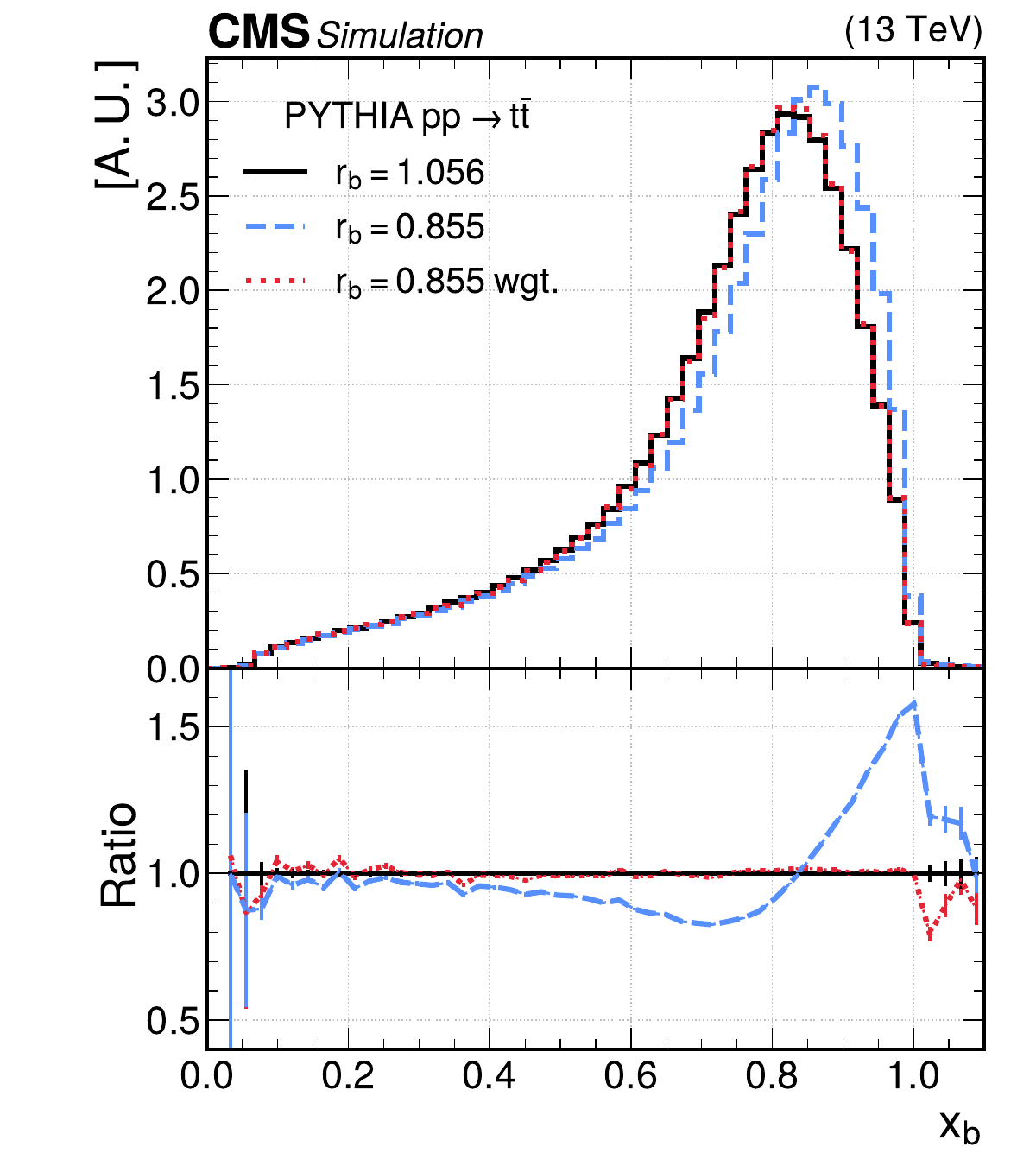}%
\hfill%
\includegraphics[width=\cmsFigWidth]{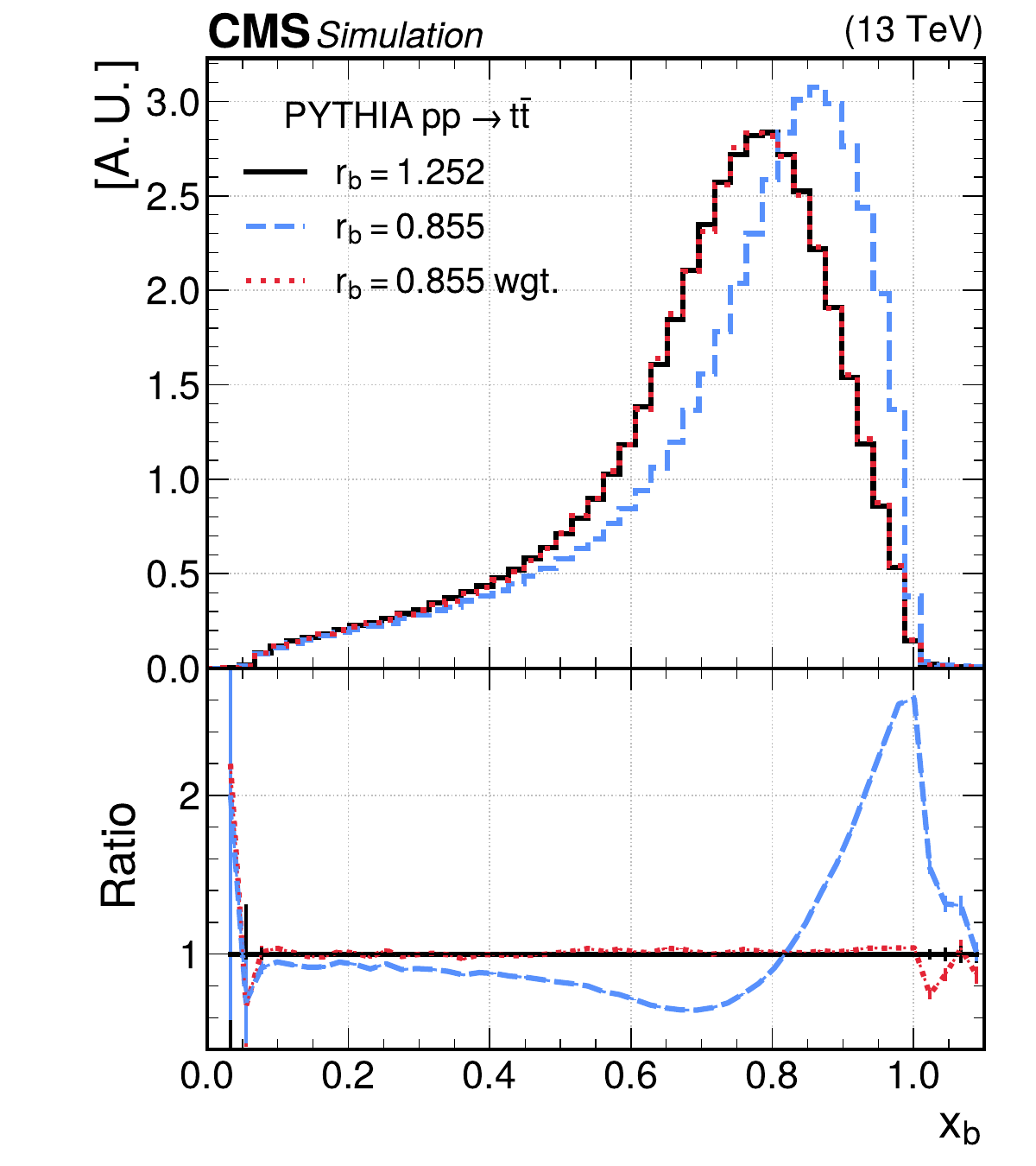} \\
\includegraphics[width=\cmsFigWidth]{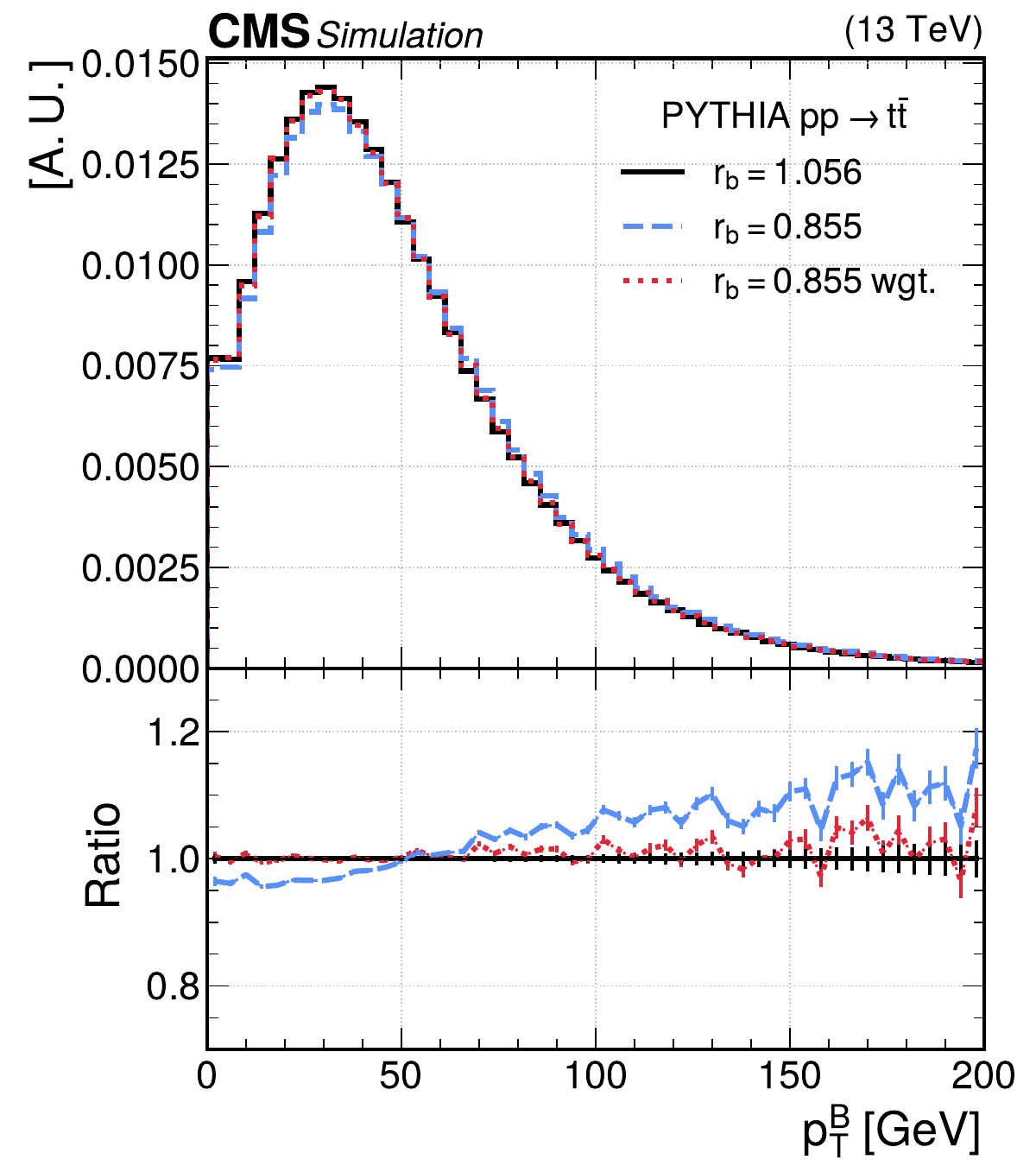}%
\hfill%
\includegraphics[width=\cmsFigWidth]{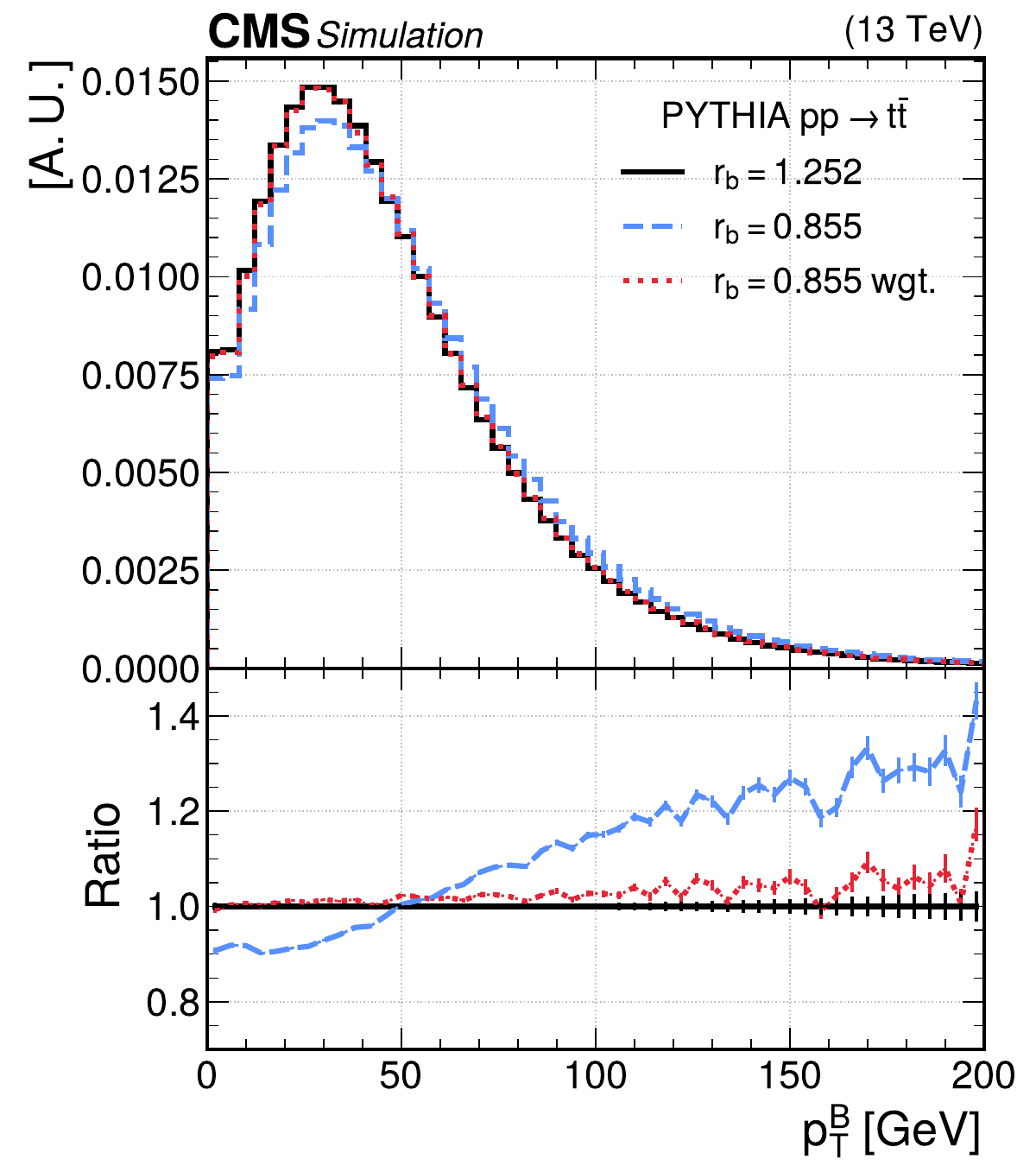}
\caption{Distributions in \xb (upper) and \ptB (lower) from \ttbar simulations with \PYTHIA8
with value $\rb=0.855$ (dashed blue line) and a second value of \rb (solid black line).
The nominal sample reweighted to $\rb=1.056$ (left) and $\rb=1.252$ (right) is shown
as red dotted lines.
Below each distribution, the ratios to the target distribution are displayed, where the vertical bars represent the statistical uncertainties.
}
\label{fig:RwgB}
\end{figure*}

The training performance is evaluated by examining the training and validation loss values,
where the NN training is stopped after about 20 epochs in both cases.
The weights obtained from the reweighting are found to be in the intervals $[0.6,1.4]$ for the
reweighting to $\rb=1.056$ and $[0.2,2.5]$ for the reweighting to $\rb=1.252$.
The performance of the reweighting is studied with statistically independent samples
with $\sim$2\,000\,000 events.
In particular, the effect of the reweighting on the \xb and \ptB distributions
is shown in Fig.~\ref{fig:RwgB}. We find that the reweighted distributions and the distributions
simulated with the target \rb values agree within the statistical uncertainties.

\begin{figure}[!ht]
\centering
\includegraphics[width=\cmsFigWidth]{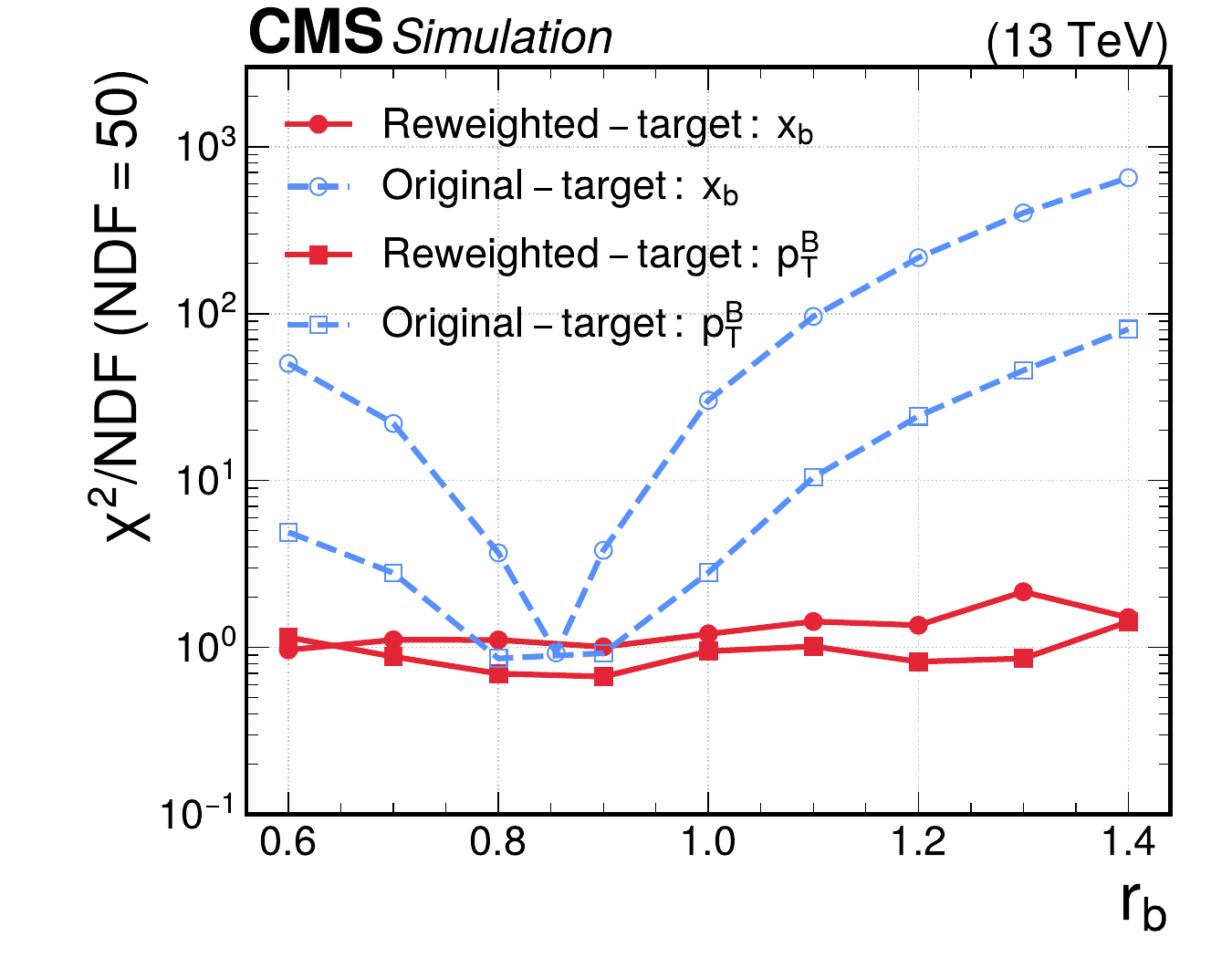}
\caption{Values of $\chisq/\ndf$ obtained for distributions in \xb (circles)
and \ptB (squares), where target distributions for events with different \rb values
are compared to a distribution with the nominal value of $\rb=0.855$ before the reweighting (blue dashed line) and
after the reweighting to the target value of \rb (red solid line).
The lines connecting the markers are shown for illustration purposes only.
}
\label{fig:Chi2B}
\end{figure}

The same procedure is applied to the continuous reweighting in \rb, using statistically independent samples with 250\,000 events.
To evaluate the goodness of the reweighting for the ten different \rb values involved in the training,
a \chisq test is performed and the results are shown in Fig.~\ref{fig:Chi2B}.
This test evaluates the difference between the target distribution generated with a given value of \rb
and the nominal distribution obtained for $\rb=0.855$.
Before the reweighting, \chisq values per number of degrees of freedom ($\ndf=50$)
of about 80 for \ptB and 650 for \xb are found for the distributions with the largest differences
in \rb to the nominal value.
After the reweighting, we find excellent agreement with the target distributions in \xb and \ptB
for all points considered in the interval $[0.6,1.4]$, with $\chisq/\ndf$ values of about unity.

\subsection{Statistical uncertainty of the method}

\begin{figure}[!ht]
\centering
\includegraphics[width=\cmsFigWidth]{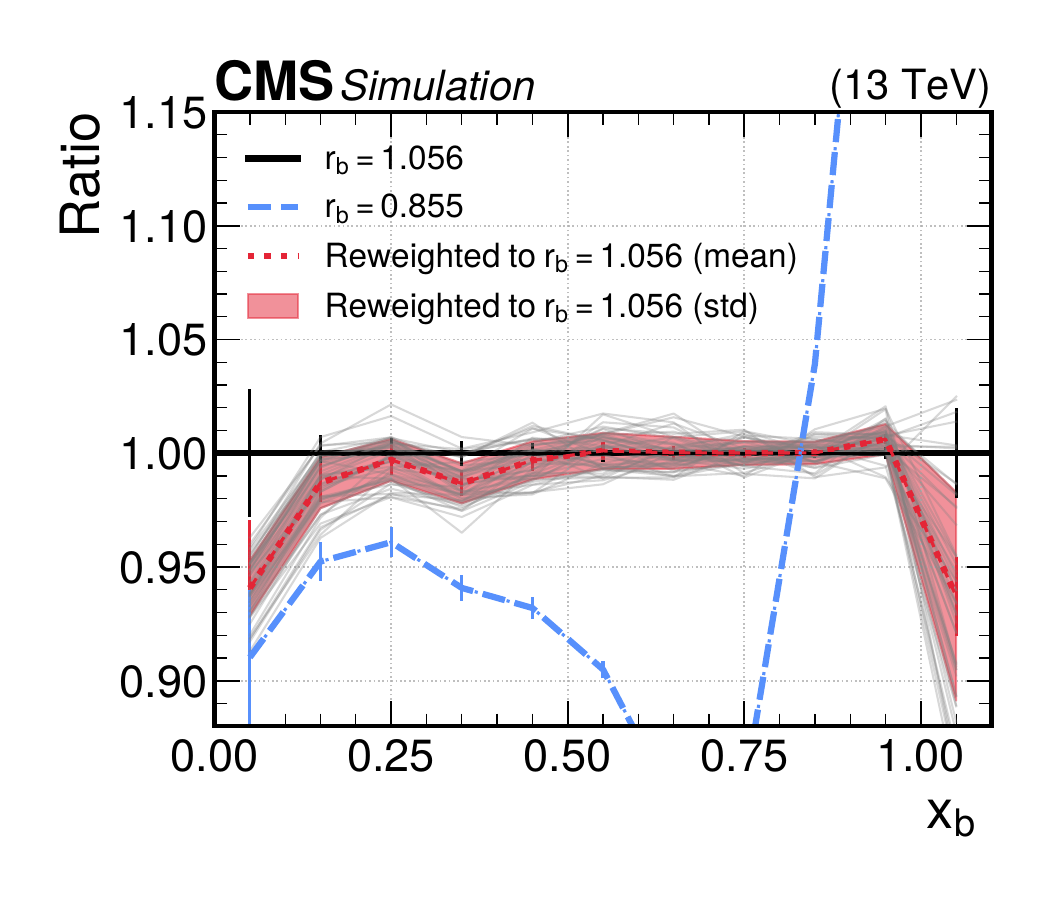}%
\hfill%
\includegraphics[width=\cmsFigWidth]{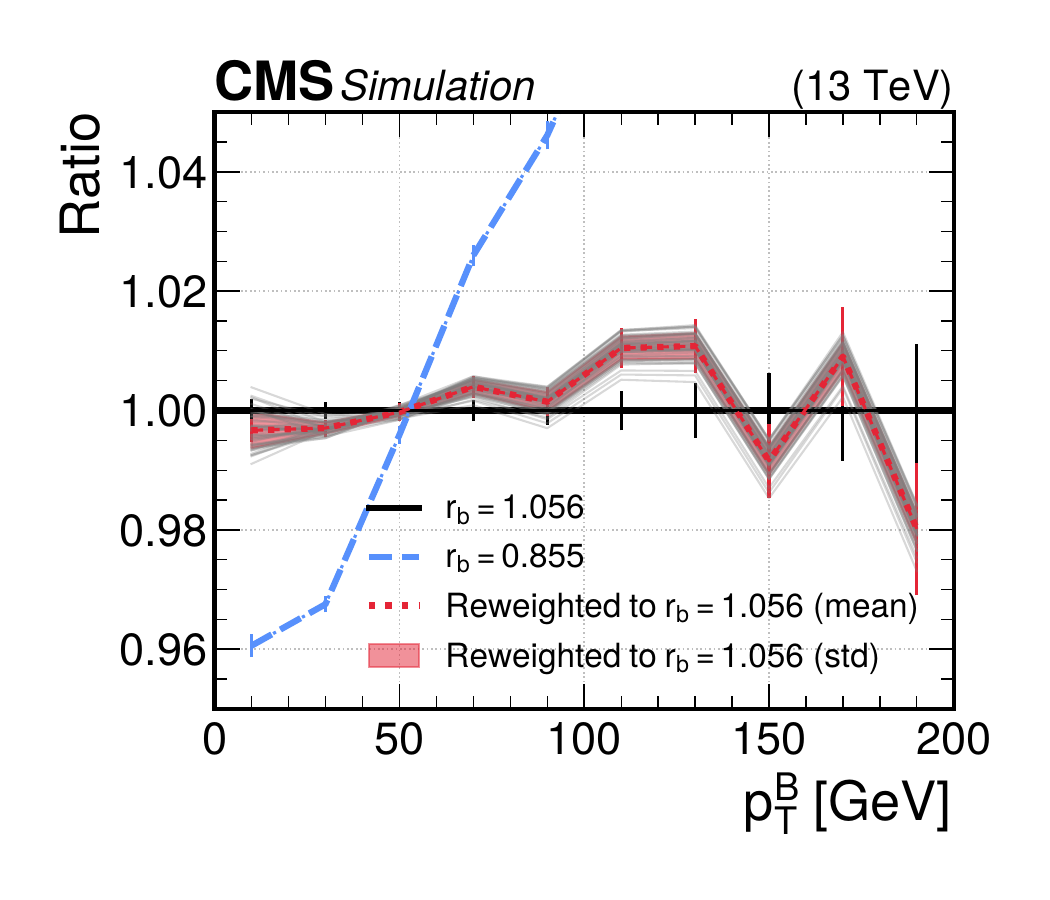}
\caption{Ratios between the \rb target distributions in \xb (\cmsLeft) and \ptB (\cmsRight),
and 50 different reweightings (grey solid lines). The ratio to the target before the reweighting
is shown as a blue dashed line and the mean of the different reweightings as a red dotted line.
The red band represents the statistical uncertainty of the method obtained from the
standard deviation of the 50 reweighted samples. The vertical bars show the statistical precision of the samples. In particular, the red bars display the average statistical uncertainty of the 50 reweighted samples.
}
\label{fig:StatBfragm}
\end{figure}

{\tolerance=800
The statistical precision of the discrete reweighting method is evaluated by
repeating the training procedure 50 times~\cite{Efron1992}.
Each training uses two million events, randomly chosen from a sample of 26 million events.
The results from the 50 different trainings, expressed in ratios of reweighted to target distributions
in \xb and \ptB, are presented in Fig.~\ref{fig:StatBfragm}.
The average from the 50 reweighted distributions is shown together with the statistical uncertainty,
obtained from the standard deviation of these distributions.
The statistical uncertainty is a few percent in both distributions.
We observe a larger statistical uncertainty of up to about 5\% for $\xb>1$.
The target sample, generated with $\rb=1.056$, and the average value of the 50 reweighted
samples are found to be compatible within the statistical uncertainty of the method,
which is of the same order as the deviation between the
target distribution and the nominal result in Fig.~\ref{fig:RwgB}.
\par}

\section{Reweighting to higher-order calculations}
\label{sec:MiNNLO}

{\tolerance=800
Recent theoretical advancements have led to very accurate predictions for \ttbar production in \pp collisions
through fixed-order computations up to NNLO~\cite{Barnreuther:2012wtj, Czakon:2012zr, Czakon:2012pz, Czakon:2013goa, Czakon:2015owf, Czakon:2016ckf,
Catani:2019iny, Catani:2019hip, Catani:2020tko}
accuracy in QCD.
These calculations rely on a power expansion in the strong coupling constant and can describe
measurements of \ttbar production remarkably well~\cite{CMS:2016oae, CMS:2017iqf, CMS:2017xio, CMS:2018tdx,
CMS:2018adi, CMS:2018tye, CMS:2019esx, CMS:2021vhb, ATLAS:2015lsn, ATLAS:2016pal, ATLAS:2016pbv, ATLAS:2017cez,
ATLAS:2017dhr, ATLAS:2019hxz, ATLAS:2019hau}.
However, in specific kinematic regimes,
an all-order resummation of radiative corrections is needed for reliable perturbative
predictions~\cite{Beneke:2011mq, Beneke:2012wb, Zhu:2012ts, Li:2013mia, Catani:2014qha, Catani:2018mei,
Ju:2020otc, Alioli:2021ggd}.
The PS simulation allows for the inclusion of soft and
collinear QCD emissions to all orders in perturbation theory, and resums the large
logarithmic corrections.
The development of PS algorithms for NLO calculations
has resulted in a leap in theoretical accuracy for describing data from the LHC,
compared to the previously available leading-order calculations matched to PS simulations.
Significant progress has been made recently to combine
NNLO calculations with PS MC generators~\cite{Hamilton:2012rf, Alioli:2013hqa, Hoche:2014uhw}.
However, \ttbar production poses a significant challenge in this respect because of the presence
of coloured particles in the initial and final state already at leading order.
The \MINNLOPS method~\cite{Monni:2019whf, Monni:2020nks, Mazzitelli:2020jio, Mazzitelli:2021mmm}
has been developed specifically for the simulation of \ttbar production
at NNLO+PS accuracy.
This method poses a significant computational complexity and cost for simulating events.
The free parameters of the corresponding MC event generator need to be determined,
and the results have to be experimentally validated before
large-scale simulations can be considered for use in experimental analyses.
\par}

In the meantime, corrections for NLO+PS simulations have been derived from a comparison to
NNLO calculations. These corrections have been obtained as a function of a single observable,
such as the \pt of the \PQt or \PAQt quarks. While these corrections can help to improve the
NLO+PS calculations in describing a single distribution~\cite{CMS:2021iuw, CMS:2021mku, ATLAS:2022xfj},
these corrections fail in capturing
the intricate higher-order corrections in the full phase space of \ttbar production.
We present a reweighting of NLO+PS simulations to NNLO+PS predictions with the DCTR method,
which captures the full kinematics of the \ttbar system.

\subsection{Network training}

We perform the reweighting of NLO+PS to NNLO+PS calculations by training a deep NN classifier to distinguish
between these two classes of simulations. The event weight for the reweighting is then
obtained with the likelihood ratio trick.
For this task, 10 million simulated \ttbar events at NLO accuracy and 10 million at NNLO accuracy
are used, obtained with the \HVQ~\cite{Frixione:2007nw} and \MINNLO~\cite{Mazzitelli:2020jio, Mazzitelli:2021mmm}
package of the \POWHEG~\cite{Frixione:2007vw, Alioli:2010xd} event generator,
both interfaced with \PYTHIA8~\cite{Sjostrand:2014zea} for the PS.
The training data are randomly split into a training set with 75\% of the events
and a validation set with 25\%.
Events after the PS are used for the NN training.
The input variables to the network are \pt, $y$, $\phi$, $m$ and the particle type
of the top quark and antiquark just before the decay, obtained from the last copy in the
\PYTHIA8 event record, and the combined \ttbar system for each event.
All input variables are standardized to take values of the same magnitude. This is achieved by using
the expectation values and standard deviations of the corresponding distributions obtained from the NLO simulation.
In the case of \pt, the logarithm of \pt is used to reduce the skewness of the input distribution.
To this end, we only reweight events based on the kinematics of the \ttbar system,
inclusive over the additional matrix element and PS radiation.

The divergences from infrared singularities are handled in the \POWHEG method by cancelling soft and
collinear real emissions by corresponding virtual corrections. This cancellation leads to a fraction of
events with negative weight, which give a negative contribution to physical observables. The fraction
of such events with negative weight amounts to about 1\% for NLO accuracy and about 10\% for NNLO accuracy.
However, the binary cross-entropy given by Eq.~\eqref{eq:L1w}, can become negatively
unbounded for negative event weights,
making the classification task potentially impossible~\cite{Nachman:2020fff}.
In such cases, the loss function can increase without bound, especially when the predicted probabilities are highly
confident but incorrect. For example, if the true classification of an event is $t_i=0$,
but the predicted class is given by $f(x_i)=1$, then Eq.~\eqref{eq:L1w} becomes
proportional to $-\wMCi\log(0) = +\abs{\wMCi}\log(0) \to -\infty$.
This unbounded growth can lead to extreme values in the loss function,
causing numerical instability during the model training process.
Consequently, the model may fail to converge to a meaningful solution,
rendering the classification task unsolvable.
This effect can be mitigated by using a large batch size, which reduces the risk of a single
event dominating the loss function. This approach works for NLO simulations, which have a
small fraction of negative events. For NNLO calculations with a significant fraction of negative
events, the training becomes unstable with the binary cross entropy.
A possible solution is the use of the mean square error (MSE) as loss function,
\begin{equation}
    \text{loss}\big(f(x)\big) = -\frac{1}{N}\sum_i^N\wMCi\big(f(x_i)-t_i\big)^2.
\end{equation}
In the case of the MSE, negative event weights still result in a reduction of the loss for
wrongly classified events. 
However, the contribution from these events is proportional to $-\wMCi$ without 
a large factor, such that the sum over sufficiently many events will result 
in a total loss function which is positive and not unbounded anymore, stabilizing the
training even when including negative events.

A batch size of $2^{17}=131\,072$ events is found to give the best results for MSE.
We reduce the learning rate of 0.001 by 60\% when no improvement is achieved after
40\% of the training epochs have been processed.

\subsection{Results}

\begin{figure}[!ht]
\centering
\includegraphics[width=\cmsFigWidth]{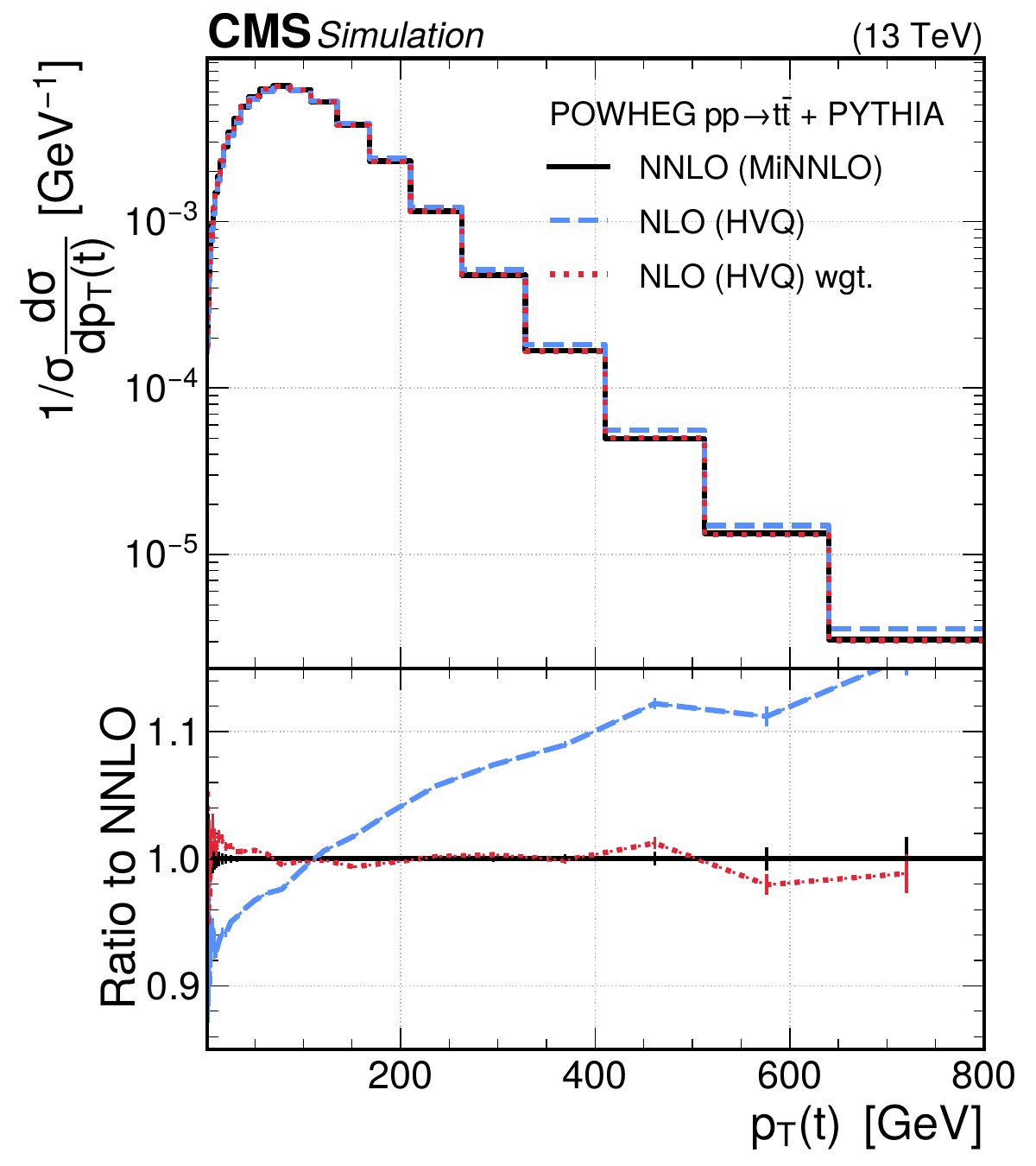}%
\hfill%
\includegraphics[width=\cmsFigWidth]{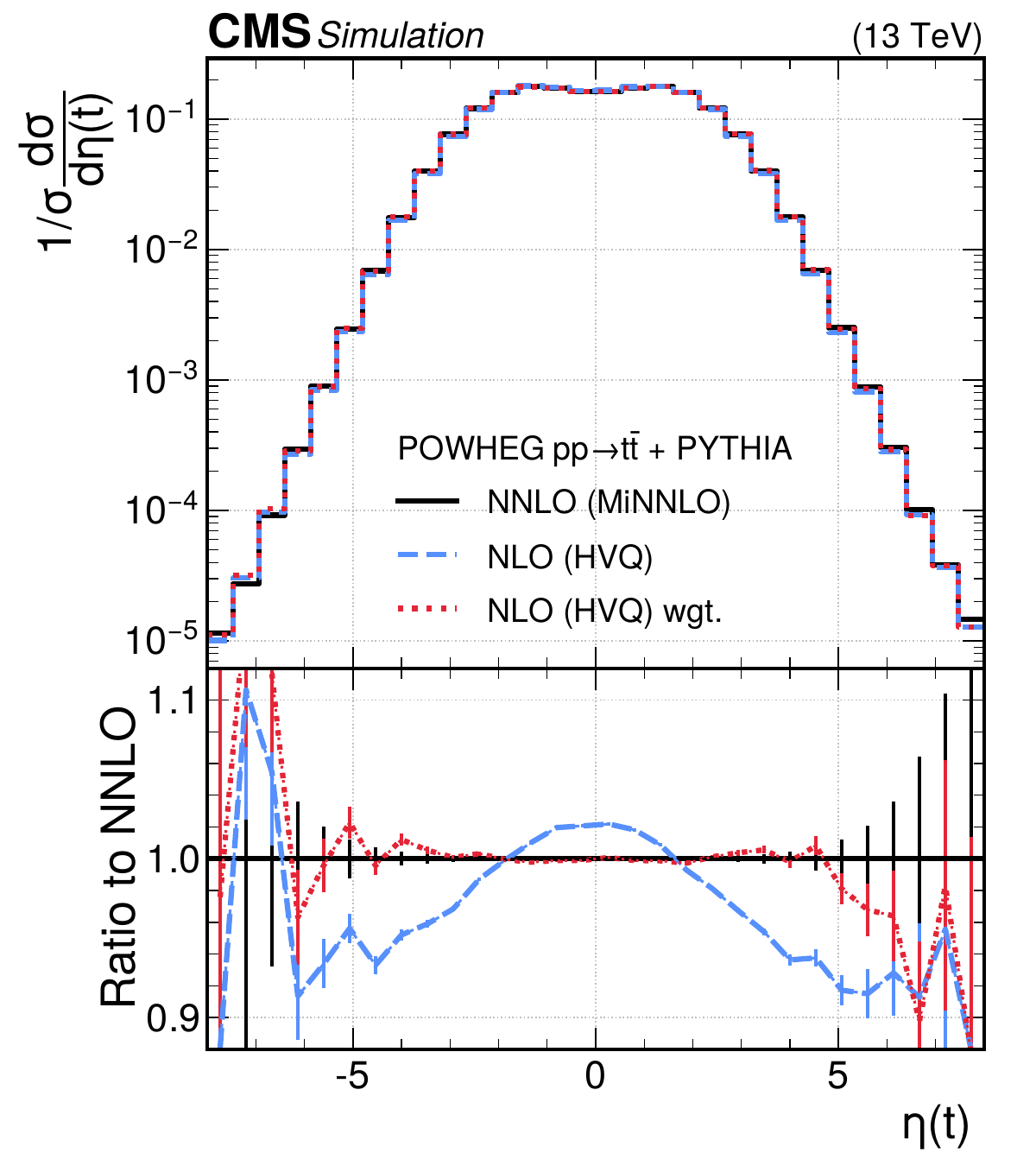}
\caption{Distributions in top quark \pt (\cmsLeft) and $\eta$ (\cmsRight)
obtained from simulations at NNLO accuracy (black solid lines),
NLO accuracy (blue dashed lines), and NLO reweighted to NNLO with the DCTR method
(red dotted lines). The ratio to the NNLO predictions is shown in the lower panels,
where the vertical bars correspond to the statistical uncertainties.
}
\label{fig:MiNNLO_t}
\end{figure}

The results from the NLO-to-NNLO reweighting are presented for about 10 million NLO events and 10 million NNLO events,
different from the ones used for the training.
Distributions in the kinematic properties of the \ttbar system are compared, treating the \PQt
and \PAQt as stable particles. In Fig.~\ref{fig:MiNNLO_t}, the distributions in
\pt and $\eta$ of the \PQt are shown. The differences between
the NLO and NNLO calculations amount to less than 10\% at low \pt, increasing to 15\% at high \pt.
The differences between the distributions in $\eta$ are smaller between the NLO and NNLO calculations
and range from 2\% in the central region to about 6\% at $\abs{\eta}=5$.
When comparing the reweighted NLO samples with the NNLO samples, we find agreement within
the statistical uncertainties.
We observe a very similar picture in the distributions of the \PAQt, which are not shown here.

\begin{figure*}[!tp]
\centering
\includegraphics[width=\cmsFigWidth]{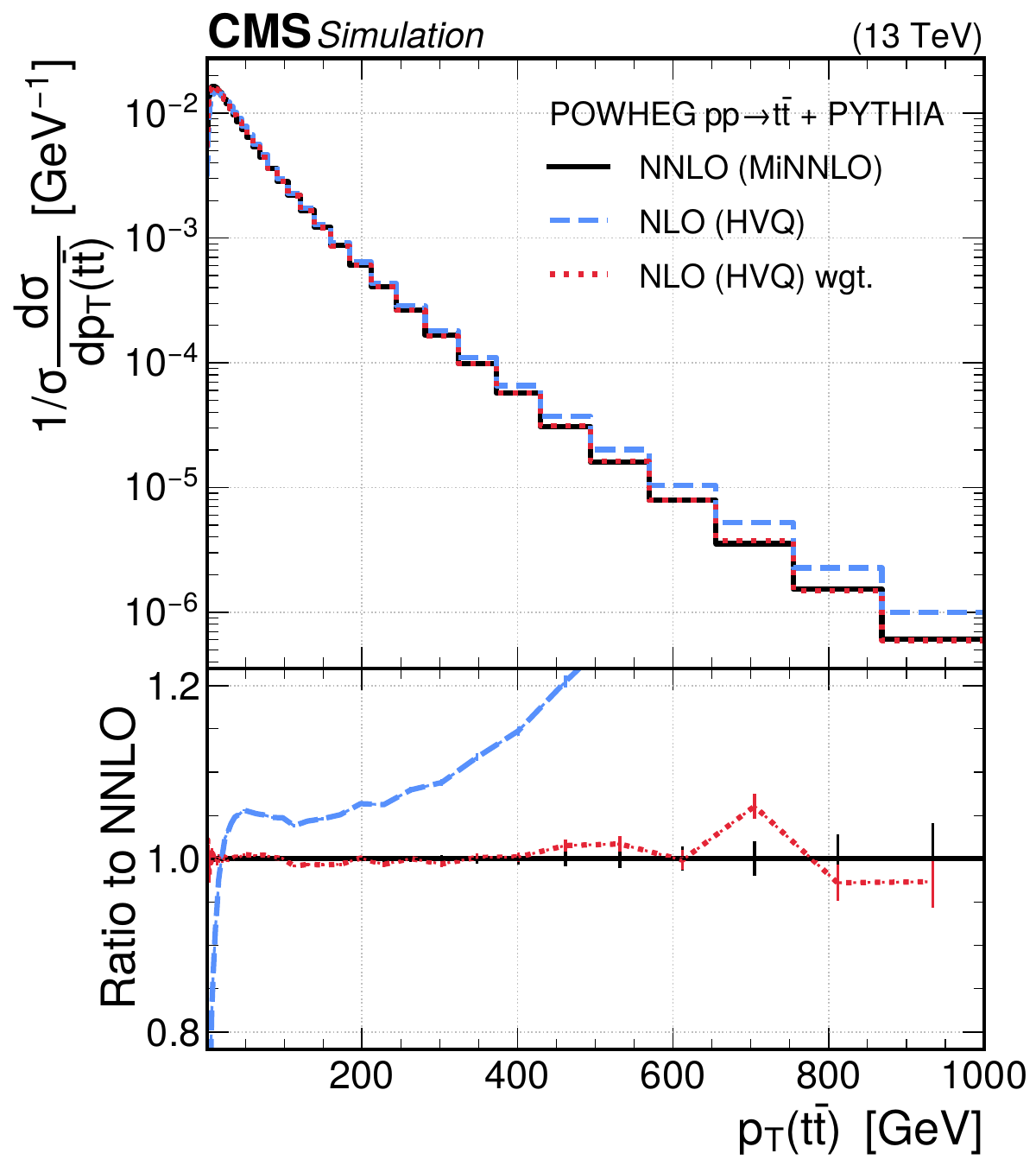}%
\hfill%
\includegraphics[width=\cmsFigWidth]{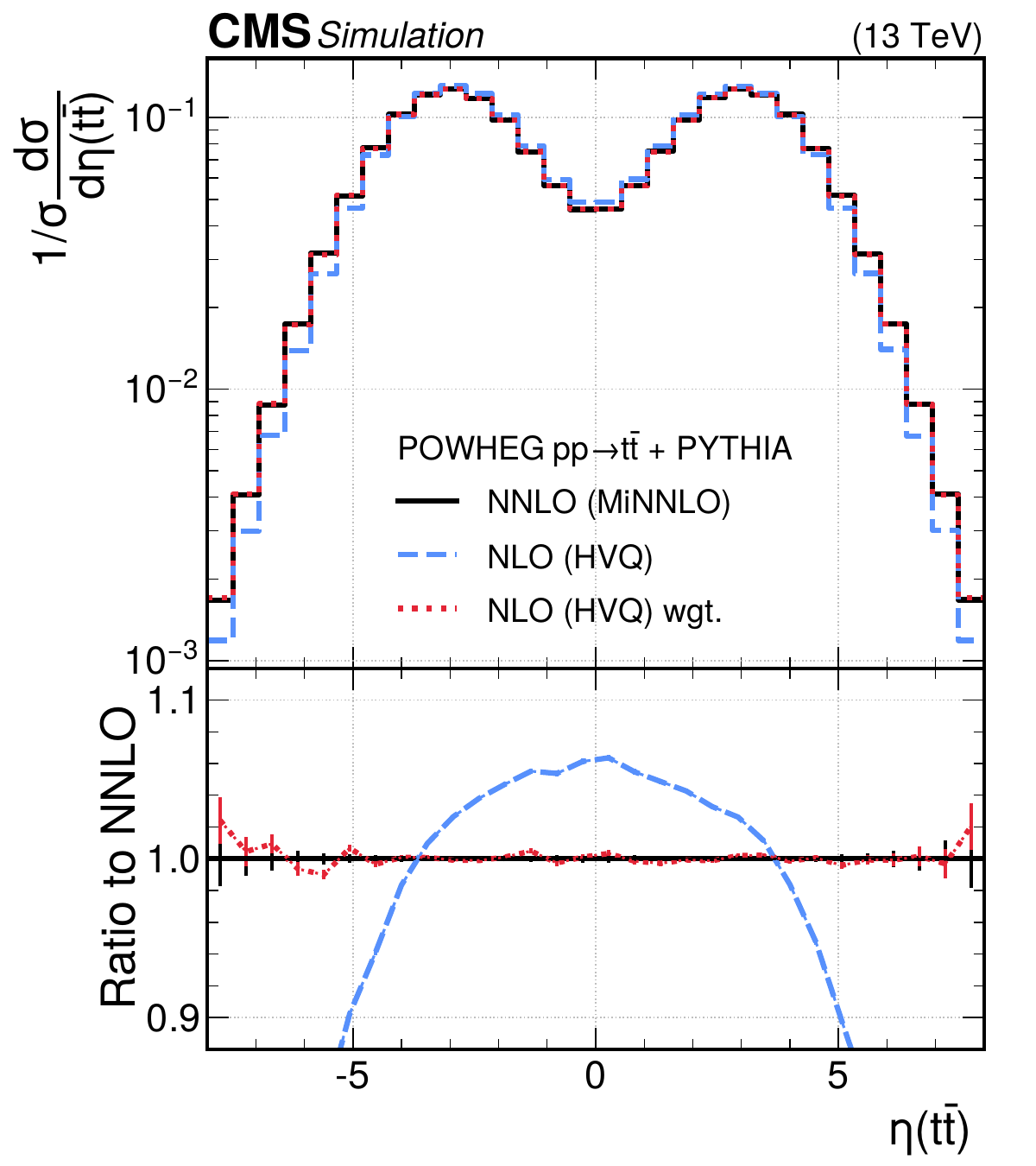} \\
\includegraphics[width=\cmsFigWidth]{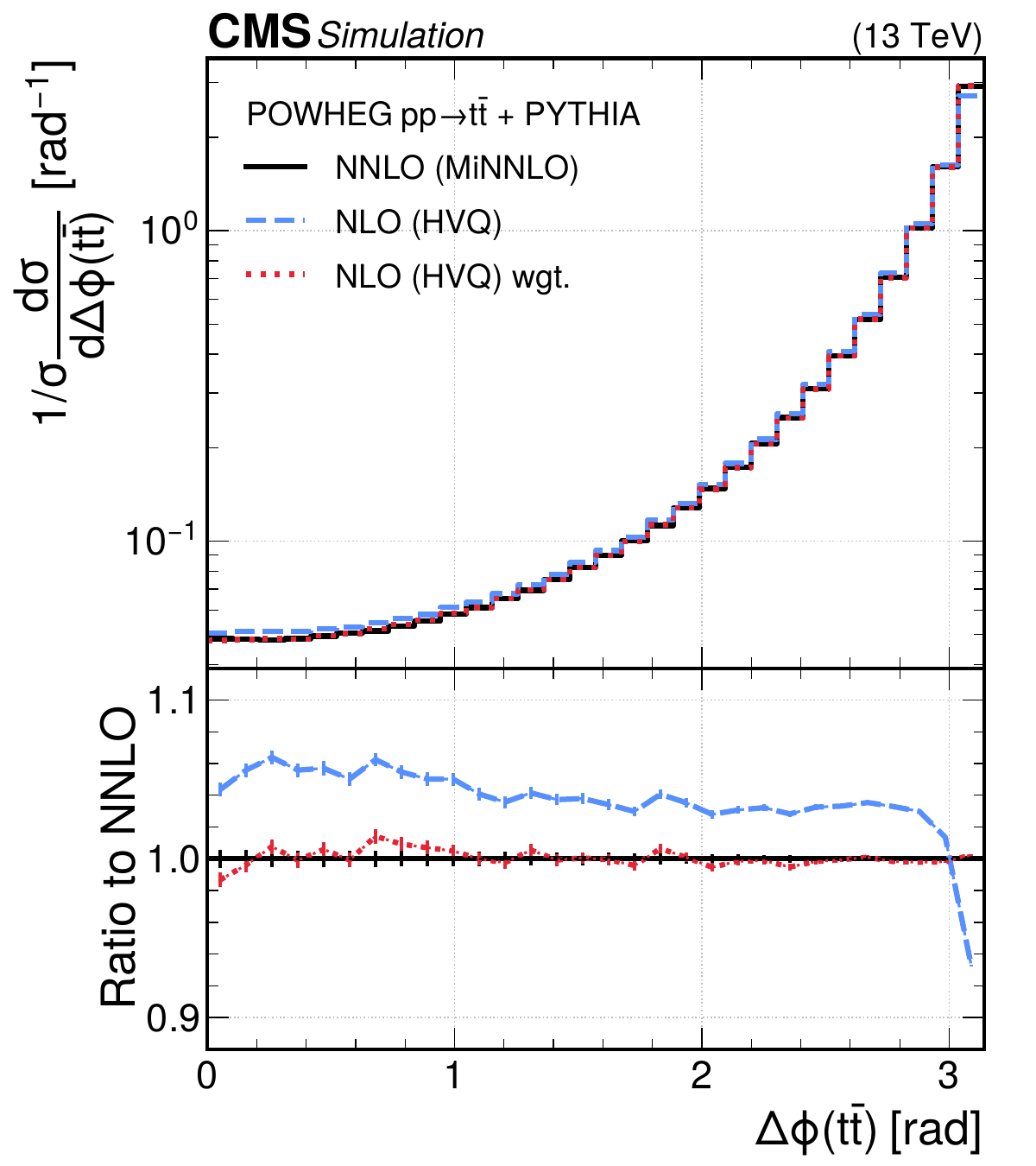}%
\hfill%
\includegraphics[width=\cmsFigWidth]{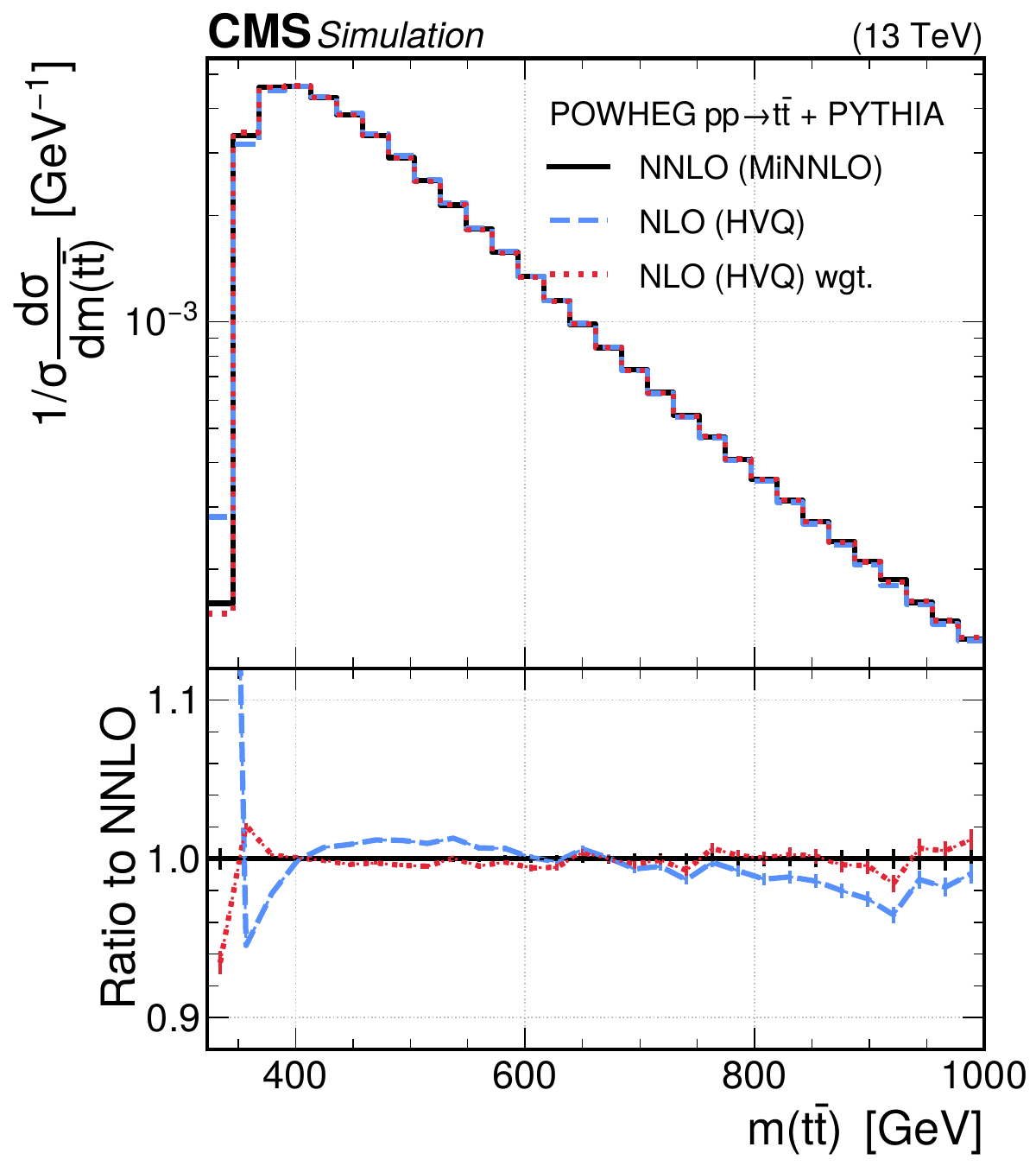}
\caption{Distributions in \pt (upper left), $\eta$ (upper right), $\Delta\phi$ (lower left),
and mass (lower right) of the \ttbar system obtained from simulations at NNLO accuracy (black solid lines),
NLO accuracy (blue dashed lines), and NLO reweighted to NNLO with the DCTR method
(red dotted lines). The ratio to the NNLO predictions is shown in the lower panels,
where the vertical bars correspond to the statistical uncertainties.
}
\label{fig:MiNNLO_tt}
\end{figure*}

{\tolerance=800
Comparisons between the NLO, NNLO, and NLO-to-NNLO reweighted
predictions for the \ttbar system are shown in Fig.~\ref{fig:MiNNLO_tt}.
In all distributions, there are significant differences between
the NLO and NNLO predictions. The NLO-to-NNLO reweighting brings the NLO predictions
into agreement with the NNLO predictions within the statistical uncertainties
over the full kinematic range of the \ttbar system.
We note that the reweighting works even for observables that were not part of the training,
such as \etatt (upper right of Fig.~\ref{fig:MiNNLO_tt})
or $\Delta\phi$, which is the difference in
azimuthal angle between the \PQt and \PAQt (lower left of Fig.~\ref{fig:MiNNLO_tt}).
The mass of the \ttbar system (lower right of Fig.~\ref{fig:MiNNLO_tt}) is also in good
agreement between the NNLO and NLO-to-NNLO reweighted predictions, which is important in
searches for beyond-the-SM effects. The DCTR reweighting achieves this simultaneously
with all other kinematic distributions of the \ttbar system, which is not possible
with traditional reweighting methods.
A limitation of the method is that additional radiation beyond the \pt-leading jet is
not accurately reweighted to the NNLO predictions, because the recoil from the \ttbar system is
distributed evenly over the remaining jets in the event.
As a result, observables related to additional radiation such as the jet multiplicity 
are not accurately reweighted to the NNLO prediction.
\par}

\begin{figure}[!ht]
\centering
\includegraphics[width=\cmsFigWidth]{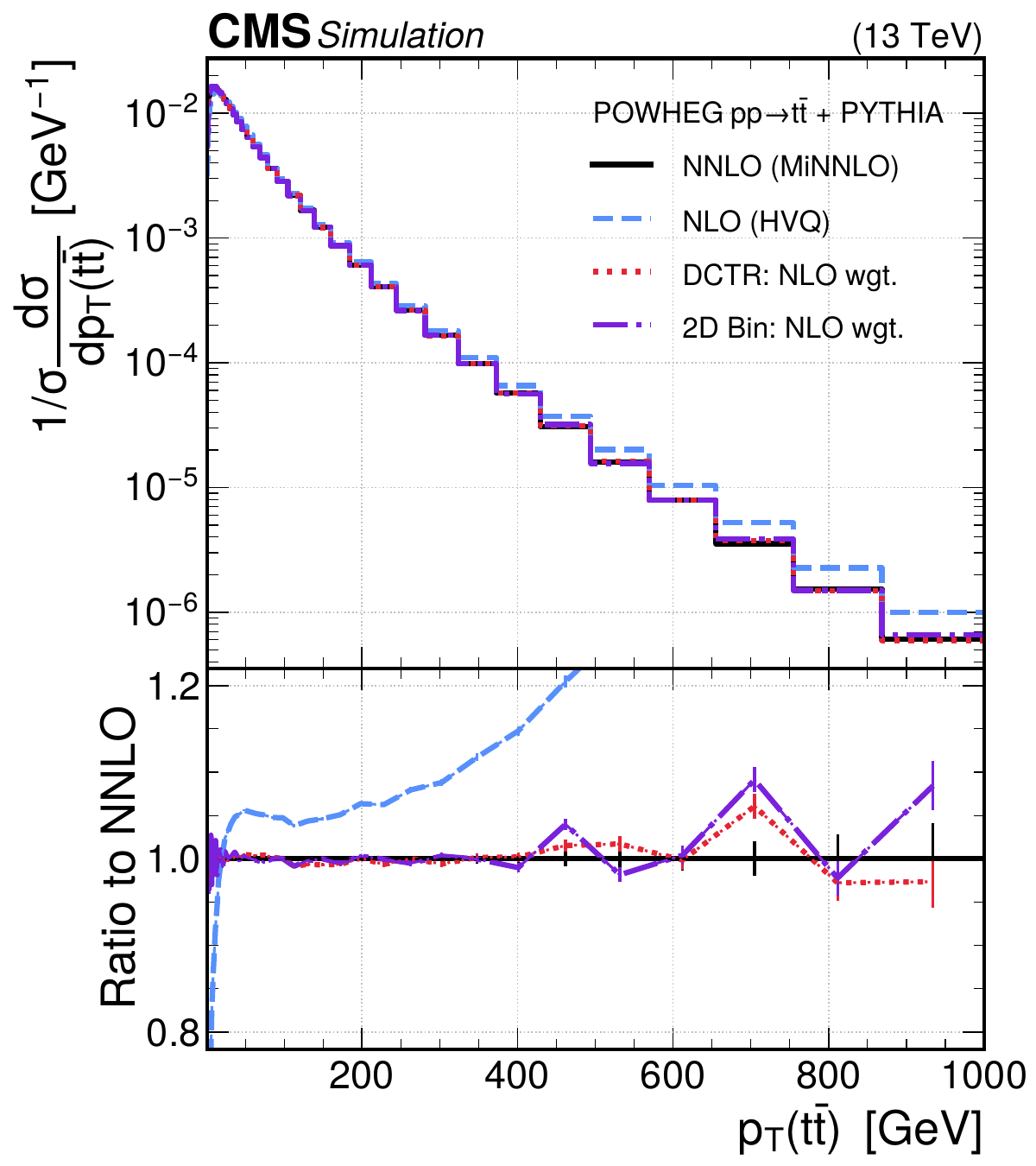}%
\hfill%
\includegraphics[width=\cmsFigWidth]{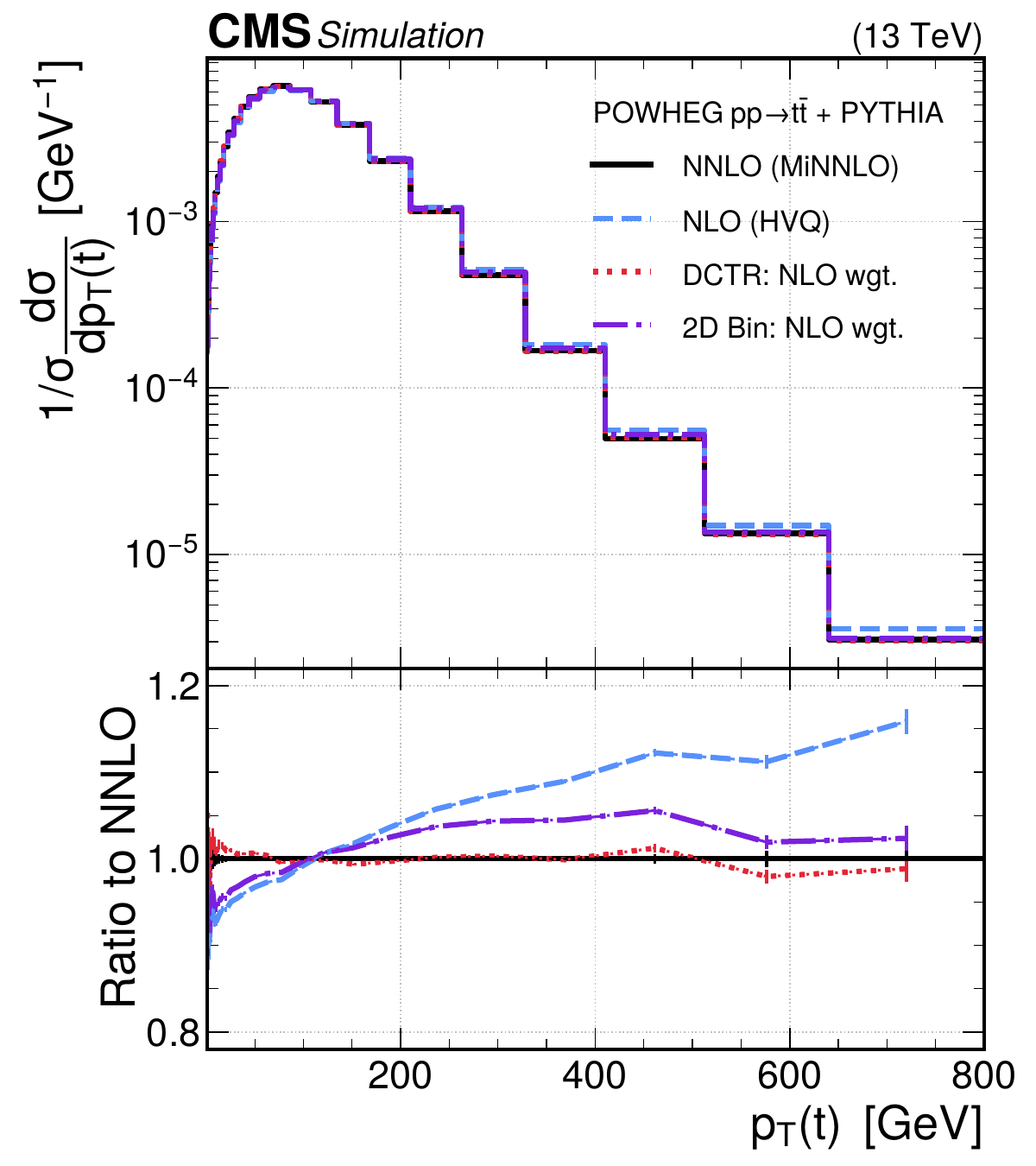}
\caption{Distributions in \pt of the \ttbar system (\cmsLeft) and \pt of the \PQt (\cmsRight)
obtained from simulations at NNLO accuracy (black solid lines),
NLO accuracy (blue dashed lines), NLO reweighted to NNLO with the DCTR method
(red dotted lines), and NLO reweighted using a two-dimensional reweighting in \pt of the \ttbar system and
of the  \PQt (violet dash-dotted line).
The ratio to the NNLO predictions is shown in the lower panels,
where the vertical bars correspond to the statistical uncertainties.
}
\label{fig:MiNNLO_2Dcomp}
\end{figure}

We provide a comparison of the DCTR reweighting with a two-dimensional (2D) reweighting
in Fig.~\ref{fig:MiNNLO_2Dcomp}. The 2D reweighting has been obtained for \pttt and \etatt,
where the ratio of NNLO to NLO predictions has been calculated
for sufficiently large bins in these two observables to reduce statistical fluctuations.
For the distribution in \pttt, as shown in Fig.~\ref{fig:MiNNLO_2Dcomp} (\cmsLeft), the 2D reweighting
works reasonably well over the full range in \pttt, where the NLO prediction is brought to agreement
with the NNLO prediction within about 2\%.
However, in the \pt of the \PQt, not used in the 2D reweighting and shown in Fig.~\ref{fig:MiNNLO_2Dcomp} (\cmsRight),
the 2D reweighting method is not entirely successful. Although the 2D-reweighted simulation
brings the NLO prediction closer to the NNLO prediction, there are differences of up to 6\%.
In contrast, the DCTR reweighting of the NLO sample accurately reproduces the NNLO prediction.
The DCTR NLO-to-NNLO reweighting can be used to correct the kinematics of the \ttbar system
for higher-order effects in the full phase space of \ttbar production. The reweighting will
have a considerable impact on ongoing and future analyses, until full NNLO+PS simulations such as
\MINNLOPS can be produced in very large samples including hadronization effects, particle decays,
and detector simulation.

\section{Implementation in CMSSW}
\label{sec:CMSSW}

An important aspect of the studies presented in this article is the availability of the
trained DCTR models within the central CMS software framework (CMSSW)~\cite{cmssw}.
The goal of the implementation is to offer all CMSSW users the ability to integrate the
trained NN models into physics analyses for the fast evaluation of systematic uncertainties
or the impact of higher-order corrections.
This encompasses the efficient computation of event weights, specific to each analysis.
The computation of these weights can be integrated either at the level of centrally produced
data formats or at any chosen stage within the data analysis.

{\tolerance=800
To enhance user flexibility, NN models are stored in the universal
open NN exchange (ONNX) format~\cite{onnx}. The ONNX program is an open-source framework,
specifically designed to facilitate interoperability and portability between different
deep-learning frameworks and tools. Major frameworks like \textsc{TensorFlow}~\cite{tensorflow2015-whitepaper},
\textsc{PyTorch}~\cite{pytorch}, and \textsc{XGBoost}~\cite{xgboost}
support conversion to and from the ONNX format, enabling seamless integration.
The runtime of ONNX for inference with trained models is compatible with CMSSW for GPU-based inference.
The CMSSW software includes already the required ONNX libraries, eliminating the
need for an installation by the user.
\par}

In general, the user has to implement the following steps to apply the DCTR reweighting
within the CMSSW software.
{\tolerance=800\begin{itemize}
    \item Read the nominal sample, \ie the sample that should be reweighted.
    \item Load the trained ONNX model.
    \item Read or calculate the inputs, according to the reweighting scenario.
    \item Standardize the inputs, identical to what has been done for the training, and pass them to the ONNX model.
    \item Read out and apply the weights to the nominal sample to obtain the requested variation.
\end{itemize}}
One significant advantage of this workflow lies in the flexibility of weight computation
and application throughout various stages of the analysis pipeline. This flexibility enables
a reweighting of events at the parton level, as well as the reweighting of events including the detector simulation.
Notably, in the latter scenario, the detector simulation has been done only once for the nominal sample
which yields a substantial reduction of computational resources, as variations of \hdamp and
\PQb quark fragmentation parameters merely require reweighting.

\section{Summary and conclusions}
\label{sec:conclsuions}

Particle physics relies on the simulation of events using Monte Carlo (MC) event generators for
data-to-theory comparisons. Data analyses require the production of several samples simulating the
same physical process to estimate systematic uncertainties or the impact of higher-order calculations.
To provide statistically significant predictions, these samples have to be very large with billions of events
generated and simulated at a high computational cost.
Nevertheless, the statistical precision from the finite size of these samples can become a limiting
factor in precision analyses. The production of sufficiently large MC samples, such that the statistical
precision of these samples is better than the statistical precision of the data, will become increasingly
prohibitive at the High-Luminosity LHC (HL-LHC) with the expected computing resources.

In this article, the method ``deep neural network using classification for tuning and reweighting (DCTR)''
has been introduced to reweight MC samples used in CMS analyses. The weights calculated with the DCTR
model enable the modification of one nominal sample to resemble other samples obtained with different parameters or
different simulation programs. This methodology avoids the need for simulating the detector response
for multiple samples by incorporating the relevant variations in a single sample.
While dedicated samples have to be generated for the training and validation of the model, these do not
need the full detector simulation and reconstruction, saving up to 75\% of the typical CPU
resources needed for the production of MC samples in CMS.
In addition, after the training of the DCTR model,
the training samples can be deleted saving storage space for several billions of events.

The DCTR method has been shown to work reliably for two important sources of modelling uncertainties in the
simulation of top quark pair (\ttbar) production. Currently, the systematic uncertainty connected to the matching of
radiation from matrix elements and the parton shower has to be estimated with dedicated samples.
The reweighting of variations in the \PQb quark fragmentation shows that a continuous reweighting in a
model parameter is possible, paving the way for the determination of model parameters directly from collision data.
Additionally, the method has been extended to reweight an NLO simulation to an NNLO one for \ttbar production,
which will allow for a fast evaluation of the impact of higher-order corrections on data analyses.
The DCTR reweighting can be seamlessly integrated into CMS analyses and is already in use by the
CMS experiment. A robust performance across a range of scenarios was demonstrated, making the method
promising for future applications in other areas as well.
For example, it can be extended to other systematic variations or applied to different physics fields beyond top quark studies.
It provides an elegant solution to address the computational challenges posed by the
production of large MC samples, particularly for the HL-LHC.

\begin{acknowledgments}
We congratulate our colleagues in the CERN accelerator departments for the excellent performance of the LHC and thank the technical and administrative staffs at CERN and at other CMS institutes for their contributions to the success of the CMS effort. In addition, we gratefully acknowledge the computing centres and personnel of the Worldwide LHC Computing Grid and other centres for delivering so effectively the computing infrastructure essential to our analyses. Finally, we acknowledge the enduring support for the construction and operation of the LHC, the CMS detector, and the supporting computing infrastructure provided by the following funding agencies: SC (Armenia), BMBWF and FWF (Austria); FNRS and FWO (Belgium); CNPq, CAPES, FAPERJ, FAPERGS, and FAPESP (Brazil); MES and BNSF (Bulgaria); CERN; CAS, MoST, and NSFC (China); MINCIENCIAS (Colombia); MSES and CSF (Croatia); RIF (Cyprus); SENESCYT (Ecuador); ERC PRG, RVTT3 and MoER TK202 (Estonia); Academy of Finland, MEC, and HIP (Finland); CEA and CNRS/IN2P3 (France); SRNSF (Georgia); BMBF, DFG, and HGF (Germany); GSRI (Greece); NKFIH (Hungary); DAE and DST (India); IPM (Iran); SFI (Ireland); INFN (Italy); MSIP and NRF (Republic of Korea); MES (Latvia); LMTLT (Lithuania); MOE and UM (Malaysia); BUAP, CINVESTAV, CONACYT, LNS, SEP, and UASLP-FAI (Mexico); MOS (Montenegro); MBIE (New Zealand); PAEC (Pakistan); MES and NSC (Poland); FCT (Portugal); MESTD (Serbia); MCIN/AEI and PCTI (Spain); MOSTR (Sri Lanka); Swiss Funding Agencies (Switzerland); MST (Taipei); MHESI and NSTDA (Thailand); TUBITAK and TENMAK (Turkey); NASU (Ukraine); STFC (United Kingdom); DOE and NSF (USA).

\hyphenation{Rachada-pisek} Individuals have received support from the Marie-Curie programme and the European Research Council and Horizon 2020 Grant, contract Nos.\ 675440, 724704, 752730, 758316, 765710, 824093, 101115353, 101002207, and COST Action CA16108 (European Union); the Leventis Foundation; the Alfred P.\ Sloan Foundation; the Alexander von Humboldt Foundation; the Science Committee, project no. 22rl-037 (Armenia); the Belgian Federal Science Policy Office; the Fonds pour la Formation \`a la Recherche dans l'Industrie et dans l'Agriculture (FRIA-Belgium); the F.R.S.-FNRS and FWO (Belgium) under the ``Excellence of Science -- EOS" -- be.h project n.\ 30820817; the Beijing Municipal Science \& Technology Commission, No. Z191100007219010 and Fundamental Research Funds for the Central Universities (China); the Ministry of Education, Youth and Sports (MEYS) of the Czech Republic; the Shota Rustaveli National Science Foundation, grant FR-22-985 (Georgia); the Deutsche Forschungsgemeinschaft (DFG), among others, under Germany's Excellence Strategy -- EXC 2121 ``Quantum Universe" -- 390833306, and under project number 400140256 - GRK2497; the Hellenic Foundation for Research and Innovation (HFRI), Project Number 2288 (Greece); the Hungarian Academy of Sciences, the New National Excellence Program - \'UNKP, the NKFIH research grants K 131991, K 133046, K 138136, K 143460, K 143477, K 146913, K 146914, K 147048, 2020-2.2.1-ED-2021-00181, and TKP2021-NKTA-64 (Hungary); the Council of Science and Industrial Research, India; ICSC -- National Research Centre for High Performance Computing, Big Data and Quantum Computing and FAIR -- Future Artificial Intelligence Research, funded by the NextGenerationEU program (Italy); the Latvian Council of Science; the Ministry of Education and Science, project no. 2022/WK/14, and the National Science Center, contracts Opus 2021/41/B/ST2/01369 and 2021/43/B/ST2/01552 (Poland); the Funda\c{c}\~ao para a Ci\^encia e a Tecnologia, grant CEECIND/01334/2018 (Portugal); the National Priorities Research Program by Qatar National Research Fund; MCIN/AEI/10.13039/501100011033, ERDF ``a way of making Europe", and the Programa Estatal de Fomento de la Investigaci{\'o}n Cient{\'i}fica y T{\'e}cnica de Excelencia Mar\'{\i}a de Maeztu, grant MDM-2017-0765 and Programa Severo Ochoa del Principado de Asturias (Spain); the Chulalongkorn Academic into Its 2nd Century Project Advancement Project, and the National Science, Research and Innovation Fund via the Program Management Unit for Human Resources \& Institutional Development, Research and Innovation, grant B39G670016 (Thailand); the Kavli Foundation; the Nvidia Corporation; the SuperMicro Corporation; the Welch Foundation, contract C-1845; and the Weston Havens Foundation (USA).
\end{acknowledgments}

\bibliography{auto_generated}
\cleardoublepage \appendix\section{The CMS Collaboration \label{app:collab}}\begin{sloppypar}\hyphenpenalty=5000\widowpenalty=500\clubpenalty=5000
\cmsinstitute{Yerevan Physics Institute, Yerevan, Armenia}
{\tolerance=6000
A.~Hayrapetyan, A.~Tumasyan\cmsAuthorMark{1}\cmsorcid{0009-0000-0684-6742}
\par}
\cmsinstitute{Institut f\"{u}r Hochenergiephysik, Vienna, Austria}
{\tolerance=6000
W.~Adam\cmsorcid{0000-0001-9099-4341}, J.W.~Andrejkovic, L.~Benato\cmsorcid{0000-0001-5135-7489}, T.~Bergauer\cmsorcid{0000-0002-5786-0293}, S.~Chatterjee\cmsorcid{0000-0003-2660-0349}, K.~Damanakis\cmsorcid{0000-0001-5389-2872}, M.~Dragicevic\cmsorcid{0000-0003-1967-6783}, P.S.~Hussain\cmsorcid{0000-0002-4825-5278}, M.~Jeitler\cmsAuthorMark{2}\cmsorcid{0000-0002-5141-9560}, N.~Krammer\cmsorcid{0000-0002-0548-0985}, A.~Li\cmsorcid{0000-0002-4547-116X}, D.~Liko\cmsorcid{0000-0002-3380-473X}, I.~Mikulec\cmsorcid{0000-0003-0385-2746}, J.~Schieck\cmsAuthorMark{2}\cmsorcid{0000-0002-1058-8093}, R.~Sch\"{o}fbeck\cmsAuthorMark{2}\cmsorcid{0000-0002-2332-8784}, D.~Schwarz\cmsorcid{0000-0002-3821-7331}, M.~Sonawane\cmsorcid{0000-0003-0510-7010}, W.~Waltenberger\cmsorcid{0000-0002-6215-7228}, C.-E.~Wulz\cmsAuthorMark{2}\cmsorcid{0000-0001-9226-5812}
% ...
\par}
\vskip\cmsinstskip
\dag:~Deceased\\
$^{1}$Also at Yerevan State University, Yerevan, Armenia\\
$^{2}$Also at TU Wien, Vienna, Austria\\
% ...
\end{sloppypar}
\end{document}